\documentclass[11pt,a4paper]{article}
\usepackage{amsmath}
\usepackage{amsfonts}
\usepackage{color}
\usepackage{graphicx}
\graphicspath{{figures/}}
\usepackage{dcolumn}
\usepackage{bm}
\usepackage[
bookmarks,bookmarksnumbered,colorlinks=true,anchorcolor=blue,
linkcolor=blue,urlcolor=blue,citecolor=blue,
breaklinks=true]{hyperref}
\usepackage{geometry}
\usepackage{hep}
\usepackage{extarrows}
\usepackage{authblk}
\usepackage{amssymb}
\usepackage[utf8]{inputenc}
\usepackage{cite}
\usepackage{pifont}
\usepackage{tikz}
\usepackage{sansmath}
\usetikzlibrary{shadings,intersections,snakes}
\usepackage{verbatim}
\numberwithin{equation}{section}   
\usepackage{slashed}
\date{\today}

\begin{document}

\title{\bf Nonlinear effective dynamics of Brownian particle in magnetized plasma}

\author[]{Yanyan Bu \thanks{yybu@hit.edu.cn}~}
\author[]{Biye Zhang \thanks{zhangbiye@hit.edu.cn (correspondence author)}~}
\author[]{Jingbo Zhang \thanks{jinux@hit.edu.cn}}

\affil[]{\it School of Physics, Harbin Institute of Technology, Harbin 150001, China}

\maketitle

\begin{abstract}
An effective description is presented for a Brownian particle in a magnetized plasma. In order to systematically capture various corrections to linear Langevin equation, we construct effective action for the Brownian particle, to quartic order in its position. The effective action is first derived within non-equilibrium effective field theory formalism, and then confirmed via a microscopic holographic model consisting of an open string probing magnetic AdS$_5$ black brane. For practical usage, the non-Gaussian effective action is converted into Fokker-Planck type equation, which is an Euclidean analog of Schr$\ddot{\rm o}$dinger equation and describes time evolution of probability distribution for particle's position and velocity.

\end{abstract}

\newpage

\tableofcontents

\allowdisplaybreaks

\flushbottom

\section{Introduction}

Brownian motion is perhaps the simplest example of non-equilibrium phenomena, which, however, has played a profound role in the development of non-equilibrium statistical mechanics \cite{Prigogine2017}. In the simplest case, a Brownian particle moving in a thermal medium is effectively described by linear Langevin equation
\begin{equation}
M \frac{d^2 }{dt^2 }q(t)+\eta_0 \frac{ d}{dt }q(t)=\chi(t), \label{linear_Langevin}
\end{equation}
where $q(t)$ and $M$ are the position and effective mass of the Brownian particle, $\eta_0$ is damping coefficient. A Gaussian white noise $\chi(t)$ could be characterized by one- and two-point functions,
\begin{equation}
\langle\chi(t)\rangle=0,\qquad \langle\chi(t)\chi(t')\rangle=2T\eta_0\delta(t-t'), \label{noise_Gauss_white}
\end{equation}
where the coefficient in the second relation is due to fluctuation-dissipation theorem. Here, $T$ is the temperature of thermal medium.

For specific purpose, linear Langevin theory \eqref{linear_Langevin}-\eqref{noise_Gauss_white} would be recast into an alternative formalism. For instance, in order to avoid repeatedly solving stochastic equation \eqref{linear_Langevin} with (infinitely) many different samplings of noise, one could equivalently consider Fokker-Planck equation, which is a deterministic differential equation for probability distribution function $\mathcal P(q, \dot{q}, t)$, where a dot means time derivative. Moreover, for a Gaussian distribution of noise, the Langevin equation \eqref{linear_Langevin} could be reformulated as a functional integral \cite{Kamenev2011}, with the weight given by the Martin-Siggia-Rose-deDominicis-Janssen (MSRDJ) action. The functional integral formalism based on MSRDJ action makes it natural to adopt modern field theoretic methods to analyze more general stochastic processes.

Indeed, linear Langevin theory \eqref{linear_Langevin}-\eqref{noise_Gauss_white} could be generalized in a number of ways \cite{Kamenev2011,Chakrabarty:2018dov,Chakrabarty:2019qcp,Chakrabarty:2019aeu,Jana:2021niz}. First, linear Langevin equation \eqref{linear_Langevin} could be made nonlinear by adding general polynomial terms such as $f_1(q,\dot q, \chi)$, which contains self-interactions for dynamical variable $q$ and nonlinear interactions between dynamical variable $q$ and noise $\xi$. One special case is multiplicative noise, which amounts to making a replacement $\chi \to f_2(q, \dot q) \chi$ in \eqref{linear_Langevin}. Second, the noise would obey a non-Gaussian distribution, and might be coloured as well, which requires to go beyond \eqref{noise_Gauss_white}. Third, isotropy would be broken by an external field, such as in magnetized thermal medium \cite{Fukushima:2015wck,Bandyopadhyay:2021zlm}. Then, beyond linear level, dynamics of transverse and longitudinal modes (with respect to external field) would get mixed. These corrections may become relevant and/or important for more realistic systems. A natural question arises: what is a more systematic way of organizing these extensions? This will be pursued here through two complementary approaches.

In this work we search for an effective description for a Brownian particle in a magnetized plasma, with potential applications in heavy-ion collisions in mind. The main purpose is to reveal nonlinear corrections to linear Langevin theory \eqref{linear_Langevin}-\eqref{noise_Gauss_white} in a systematic way. This will be achieved by non-equilibrium effective field theory (EFT) formalism for a quantum many-body system at finite temperature \cite{Crossley:2015evo,Glorioso:2016gsa,Glorioso:2017fpd,Glorioso:2018wxw} (see \cite{Haehl:2015foa,Haehl:2015uoc,Haehl:2018lcu} for an alternative approach). Within such a formalism, dynamics of Brownian particle is entirely dictated by an effective action, which will be constructed on a set of symmetries. The effective action may be thought of as generalization of the MSRDJ action for a linear theory \eqref{linear_Langevin}-\eqref{noise_Gauss_white}. Moreover, the effective action contains ``free parameters'' representing UV physics and information of the state as well. Generically, it is challenging to compute those free parameters from an underlying UV theory (here, it is a closed system consisting of the Brownian particle and the magnetized plasma). Given that quark-gluon plasma produced in heavy-ion collisions is strongly coupled, we turn to a microscopic holographic model and derive the effective action (including values of free parameters).

Holography \cite{Maldacena:1997re,Gubser:1998bc,Witten:1998qj} is insightful in understanding symmetry principles underlying non-equilibrium effective action. Of particular importance is the dynamical Kubo-Martin-Schwinger (KMS) symmetry \cite{Crossley:2015evo,Glorioso:2016gsa,Glorioso:2017fpd} acting on dynamical variable of the effective action, which guarantees the generalized fluctuation-dissipation theorem \cite{Wang:1998wg,Hou:1998yc} at full level. In \cite{Crossley:2015evo,Glorioso:2016gsa,Glorioso:2017fpd}, dynamical KMS symmetry is implemented in the classical statistical limit where $\hbar \to 0$, which corresponds to neglecting quantum fluctuations in the effective theory. However, for a holographic theory, the mean free path is $\sim \hbar /T$, which implies that gradient expansion would generally inevitably bear quantum fluctuations \cite{deBoer:2018qqm}. Via the example of Brownian motion, we will elaborate on this point from both non-equilibrium EFT approach and holographic calculation. Intriguingly, imposition of a constant translational invariance (i.e., $q \to q + c$ with $c$ a constant) renders resultant effective theory to be of classical statistical nature.

While effective action formalism is more systematic in covering nonlinear corrections alluded above, it turns out to be inconvenient to convert non-Gaussian effective action into Langevin type equation \cite{Crossley:2015evo}. The main obstacle stems from non-Gaussianity in $a$-variable (to be defined below), which prohibits from carrying out Hubbard-Stratonovich transformation \cite{Kamenev2011}. Interestingly, we are able to convert the non-Gaussian effective action constructed in present work into Fokker-Planck type equation, which will be useful in numerical study.

The rest of this paper will be structured as follows. In section \ref{effective_theory}, we clarify the set of symmetries and construct effective action for Brownian particle, which is further put into Fokker-Planck type equation. In section \ref{study_holo_model}, we derive effective action for a Brownian particle moving in magnetized thermal plasma from a holographic perspective. In section \ref{summary} we summarize and outlook future directions. Appendices \ref{non-commutativity} and \ref{KMS_all} provide further calculational details.

\section{Effective dynamics from symmetry principle} \label{effective_theory}

\begin{figure}[htbp]
\centering
\includegraphics[width=0.7\textwidth]{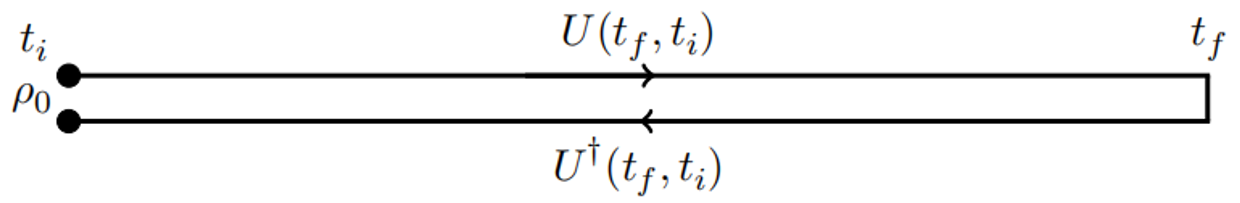}
\caption{The SK closed time contour: $\rho_0$ is initial density matrix, and $U(t_f,t_i)$ is the time-evolution operator from initial time $t_i$ to final time $t_f$.} \label{SK_contour_1}
\end{figure}

Dynamics of a closed system consisting of a Brownian particle and a thermal medium is presumably described by an action
\begin{align}
S_{\rm C}= S_{\rm p}[q] + S_{\rm th} [\Phi] + S_{\rm int}[q,\Phi], \label{action_closed}
\end{align}
where $S_{\rm p}[q]$ is action for the Brownian particle, $S_{\rm th}[\Phi]$ describes microscopic theory of the constitutes (collectively denoted as $\Phi$) for the thermal medium, and $S_{\rm int}[q,\Phi]$ is the interaction between Brownian particle and constituents of thermal medium. In principle, effective action for the Brownian particle would be obtained by integrating out degrees of freedom $\{\Phi \}$ for the thermal medium, as illustrated below:
\begin{align}
Z = \int [Dq] [D\Phi] e^{{\rm i} S_{\rm C}} = \int [D q] e^{{\rm i} I[q]},  \label{Z_Wilson_RG}
\end{align}
where $I[q]$ is the desired effective action. For such a quantum many-body system, time evolution of the state will {\it effectively} go forward and backward along the Schwinger-Keldysh (SK) closed time contour, see Figure \ref{SK_contour_1}.
Apparently, when carrying out the ``integrating out'' procedure in \eqref{Z_Wilson_RG}, one shall place the closed system on the SK closed time contour of Figure \ref{SK_contour_1}. Resultantly, the degrees of freedom are doubled, $q\to (q_1, q_2)$, where the subscripts $1,2$ correspond to the upper and lower branches of Figure \ref{SK_contour_1}.

Except for a few simple models \cite{Caldeira:1982iu,Chakrabarty:2018dov,Chakrabarty:2019qcp,Glorioso:2018wxw,Yao:2020eqy,Yao:2021lus}, it is very challenging to implement the ``integrating out'' procedure illustrated in \eqref{Z_Wilson_RG}. It is thus natural to {\it construct} the effective action based on symmetry principle, which will be pursued here.

\subsection{Construction of effective action} \label{construct_eff_action}

The effective action $I[q_1; q_2]$ is usually presented in $(r,a)$-basis:
\begin{align}
q_r \equiv \frac{1}{2}(q_1 +q_2), \qquad \qquad q_a \equiv q_1 - q_2.
\end{align}
where $q_r$ is the physical variable and $q_a$ is an auxiliary variable (conjugate to noise $\xi(t)$). We summarize various symmetries and constraints obeyed by the effective action $I[q_1; q_2] = I[q_r;q_a]$ for a Brownian particle moving in a magnetized plasma.

$\bullet$ \underline{\bf $Z_2$-reflection symmetry}

Take the complex conjugate of partition function \eqref{Z_Wilson_RG}, we find the reflection conditions,
\begin{align}
I^*[q_1;q_2]=-I[q_2;q_1] \Longleftrightarrow I^*[q_r; q_a] = - I[q_r; -q_a].
\end{align}

$\bullet$ \underline{\bf Normalization condition}

If we set the two coordinates to be the same $q_1 = q_2 = q $, we find 
\begin{align}
{\rm Tr}\left(U(+\infty,-\infty;q)\rho_0 U^\dagger(+\infty,-\infty;q)\right)={\rm Tr}\rho_0 = 1,
\end{align}
which
lead to the normalization condition,
\begin{align}
I[q_r; q_a=0]=0. \label{Z2_qa=0}
\end{align}
So, it will be convenient to present the effective action as an expansion in number of $q_a$-variable.

Moreover, for the path integral based on $I[q_r; q_a]$ to be well-defined, imaginary part of $I[q_r; q_a]$ should be non-negative:
\begin{align}
\text{Im}\left(I[q_r;q_a]\right) \geq 0, \label{imaginary_positive}
\end{align}
which will constrain some parameters in the effective action.

$\bullet$ \underline{\bf Dynamical KMS symmetry}
\begin{align}
I[q_1; q_2] = I[\tilde q_1; \tilde q_2], \label{dynamical_KMS_symmetry}
\end{align}
where
\begin{align}
\tilde q_1(-t) = q_1(t), \qquad \tilde q_2(-t) = q_2(t- {\rm i} \beta). \label{dynamical_KMS_transform}
\end{align}
Here, $\beta =1/T$ is the inverse temperature of plasma medium. The dynamical KMS symmetry is crucial in formulating an EFT for quantum many-body system at finite temperature. It guarantees the generalized nonlinear fluctuation-dissipation theorem (FDT) at full quantum level \cite{Wang:1998wg}, which originates from time-reversal invariance of underlying microscopic theory and relies on the fact that initially the system is in a thermal state. 

Intriguingly, it is possible to take classical statistical limit of dynamical KMS symmetry so that only thermal fluctuations in EFT will survive \cite{Crossley:2015evo}. Let us properly restore Planck constant
\begin{align}
\beta \to \hbar \beta, \qquad q_r \to q_r, \qquad q_a \to \hbar q_a.
\end{align}
Then, the classical statistical limit is achieved by taking $\hbar \to 0$ in the effective action. Consequently, the classical statistical limit of \eqref{dynamical_KMS_transform} becomes
\begin{align}
\tilde q_r(-t) = q_r(t), \qquad \tilde q_a(-t) = q_a(t) + {\rm i} \beta \partial_t q_r(t). \label{classical_statistical_limit}
\end{align}

On the other hand, when the mean free path is of order $\hbar \beta$ (as for a holographic theory), derivative expansion adopted in the construction of effective action would suggest considering $\hbar \beta$-expansion\footnote{Here, we keep the expansion to first order in $\beta$ for later purpose.} of the dynamical KMS transformation \eqref{dynamical_KMS_transform}:
\begin{align}
&\tilde q_r(-t) = q_r(t) - \frac{\rm i}{2}\hbar \beta \partial_t q_r(t) + \frac{\rm i} {4} \hbar^2 \beta \partial_t q_a(t), \nonumber \\
&\tilde q_a(-t) = q_a(t) + {\rm i} \beta \partial_t q_r(t) - \frac{\rm i}{2} \hbar \beta \partial_t q_a(t). \label{hbar_beta_expansion}
\end{align}
Interestingly, taking $\hbar \to 0$ in \eqref{hbar_beta_expansion} will recover the classical statistical limit \eqref{classical_statistical_limit}.

The $Z_2$-reflection symmetry and dynamical KMS symmetry are generic to non-equilibrium EFT. Specific to the problem of Brownian motion, we will impose additional symmetries.

$\bullet$ \underline{\bf $Z_2$-parity}
\begin{align}
I[q_1;q_2] = I[-q_1; -q_2] \Longleftrightarrow I[q_r;q_a] = I[-q_r; -q_a] \label{Z2_parity}
\end{align}
which means the action $I$ contains only even powers of $q_{r,a}$.

$\bullet$ \underline{\bf Rotational symmetry}
\begin{align}
I[\hat q_1; \hat q_2] = I[q_1; q_2], \qquad {\rm with} \quad \hat q_1 = \mathcal R q_1, \quad \hat q_2 = \mathcal R q_2, \label{rotatial_invariance}
\end{align}
where $\mathcal R$ denotes rotational transformation in space. Relevant to present work, the rotational symmetry will be reduced into $SO(2)_{xy}$ thanks to presence of a background magnetic field along $z$-direction.

$\bullet$ \underline{\bf Constant translational symmetry}
\begin{align}
I[q_1+c; q_2+c] = I[q_1;q_2] \Longleftrightarrow I [q_r+c; q_a] = I [q_r; q_a]  \label{constant_translation}
\end{align}
where $c$ is a constant. This symmetry is due to homogeneous property of the plasma medium. Under this symmetry, the dependence of $I[q_r;q_a]$ on particle's position will be through the velocity $\dot q_r$. Interestingly, combined with the dynamical KMS symmetry, this symmetry will stringently constrain the form of $I[q_r;q_a]$, which will accidently make the dynamical KMS symmetry at quantum level indistinguishable from its classical statistical limit, at least valid at first order in derivative expansion.

Now it is ready to present the effective action for Brownian particle in magnetized plasma. The effective action $I = \int dt L_{\rm SK}$ will be organized by employing amplitude expansion in $q_{r,a}$ and derivative expansion. Schematically, the effective Lagrangian is expanded as
\begin{align}
L_{\rm SK} = L_{\rm SK}^{(2)} + L_{\rm SK}^{(4)}+ \cdots,
\end{align}
where superscript denotes number of $q_{r,a}$-variables. Here, we have imposed the $Z_2$-parity \eqref{Z2_parity}.

First, we consider the quadratic Lagrangian $L_{\rm SK}^{(2)}$. Recall that constant translational symmetry \eqref{constant_translation} tells that effective Lagrangian should be a functional of $\dot q_r$ instead of $q_r$. After imposing the rotational symmetry \eqref{rotatial_invariance}, the most general quadratic Lagrangian is\footnote{Throughout this paper, the magnetic field is along $z$-direction, and indices $i,j$ denote transverse directions. Here, we have ignored terms like $B\epsilon_{ij} \dot q_a^i q_a^j, M^\times \epsilon_{ij} \ddot q_r^i q_a^j$ since they will not appear in our holographic model.}
\begin{align}
L_{\rm SK}^{(2)}= -M^{\rm T} q^i_a \ddot q^i_r - M^{\rm L} q^z_a \ddot q^z_r  - q^i_a \eta^{\rm T}\dot q^i_r - q^z_a \eta^{\rm L}\dot q^z_r  + \frac{\rm i }{ 2 }\left[ q^i_a \xi^{\rm T} q^i_a + q^z_a \xi^{\rm L} q^z_a \right]- Q B\epsilon_{ij}\dot q^i_r q_a^j,  \label{EFquadratic}
\end{align}
where constants $M^{\rm{T,L}}$ are identified with effective mass for transverse and longitudinal modes, respectively. $\eta^{\rm{T,L}}, \xi^{\rm{T,L}}$ are functionals of time derivative operator $\partial_t$, and could be expanded in the hydrodynamic limit
\begin{align}
\mathfrak{C}^{\rm{T,L}} = \mathfrak{C}^{\rm{T,L}}_0 + \mathfrak{C}^{\rm{T,L}}_1 \partial_t + \mathfrak{C}^{\rm{T,L}}_2 \partial_t^2 + \cdots, \qquad {\rm with} \quad  \mathfrak{C} = \eta, \xi. \label{expansion_eta_xi}
\end{align}
The $Q$-term represents the Lorentz force (with $Q$ the charge of Brownian particle), which is not a medium effect.

The requirement \eqref{imaginary_positive} sets inequality relations for $\xi^{\rm{T,L}}$, e.g., in accord with derivative expansion \eqref{expansion_eta_xi}, we have
\begin{align}
\xi^{\rm T}_0 \geq 0, \qquad  \xi^{\rm L}_0 \geq 0; \qquad \xi^{\rm T}_{2n} \leq 0, \qquad  \xi^{\rm L}_{2n} \leq 0, \qquad n=1,2,3,\cdots.
\end{align}

Finally, we turn to impose the dynamical KMS symmetry \eqref{dynamical_KMS_symmetry} at the full level \eqref{dynamical_KMS_transform}. At quadratic order, this is equivalent to the familiar FDT
\begin{align}
\eta^{\rm T}(\omega)=\coth\left(\frac{ \beta \omega}{ 2 }\right) {\rm Im} \left[ {\rm i} \omega \xi^{\rm T}(\omega) \right], \qquad
\eta^{\rm L}(\omega)=\coth\left(\frac{ \beta \omega}{ 2 }\right) {\rm Im} \left[ {\rm i} \omega \xi^{\rm L}(\omega)  \right], \label{FDT_linear}
\end{align}
where we turn to frequency domain by $\partial_t \to - {\rm i} \omega$. Since we will be interested to truncate the expansion \eqref{expansion_eta_xi} to leading order, the linear FDT becomes
\begin{align}
\eta_0^{\rm T} = \frac{1}{2} \beta \xi_0^{\rm T}, \qquad \eta^{\rm L}_0 = \frac{1}{2}\beta \xi_0^{\rm L}.
\end{align}
Through Legendre transformation \cite{Crossley:2015evo,Glorioso:2018wxw}, it is direct to show that quadratic Lagrangian $L_{\rm SK}^{(2)}$ is equivalent to a linear Langevin theory.

Next we turn to quartic Lagrangian $L_{\rm SK}^{(4)}$, which contains mixing effects between transverse and longitudinal modes. With symmetries \eqref{Z2_qa=0}, \eqref{Z2_parity}, \eqref{rotatial_invariance}, and \eqref{constant_translation} imposed, the most general quartic Lagrangian is

\begin{align}
L^{(4)}_{\rm SK}= & \kappa^{\rm T} \dot q_r^i q_a^i(q_a^j)^2 + \kappa^{\rm L} \dot q_r^z (q_a^z)^3 + \kappa^{\times}_2 \dot q_r^i q_a^i (q_a^z)^2 + \kappa^{\times}_1\dot q_r^z q_a^z (q_a^i)^2 \nonumber\\
&+\frac{\rm i }{ 4!}\left[ \zeta^{\rm T} (q_a^i)^2(q_a^j)^2 + \zeta^{\rm L}(q_a^z)^4+ \zeta^{\times} (q_a^i)^2(q_a^z)^2\right], \label{EFquartic}
\end{align}
where we have truncated at first order in derivative expansion. Here, in contrast with \eqref{EFquadratic}, various coefficients in \eqref{EFquartic} are constants.

From the constraint \eqref{imaginary_positive}, we have
\begin{align}
\zeta^{\rm T} \geq 0,\qquad\zeta^{\rm L} \geq 0,\qquad\zeta^{\times} \geq 0.  \label{imaginary_positive_quartic}
\end{align}
Finally, we impose dynamical KMS symmetry \eqref{dynamical_KMS_symmetry}. In the classical statistical limit \eqref{classical_statistical_limit}, this implies
\begin{align}
\kappa^{\rm T}=-\frac{1}{12}\beta \zeta^{\rm T},\qquad \kappa^{\rm L}=-\frac{1}{12}\beta \zeta^{\rm L}, \qquad \kappa^{\times}_1=\kappa^{\times}_2=-\frac{1}{24}\beta \zeta^{\times}.  \label{quarticKMS}
\end{align}

Intriguingly, it can be shown that if we had imposed \eqref{hbar_beta_expansion}, we would obtain the same conclusion \eqref{quarticKMS}. This seemingly implies for the Brownian motion example considered in this work, classical statistical limit \eqref{classical_statistical_limit} is indistinguishable from the high-temperature limit \eqref{hbar_beta_expansion}. This is indeed attributed to the constant translational invariance \eqref{constant_translation}. More precisely, if this symmetry is relaxed, the following terms shall be added to \eqref{EFquartic} (each with an independent coefficient)
\begin{align}
q_r q_a^3, \quad q_r^2 q_a^2, \quad q_r^3 q_a, \quad  \dot q_r q_r q_a^2, \quad \dot q_r q_r^2 q_a, \quad \cdots. \label{added_terms}
\end{align}
Then, it is straightforward to check that under classical statistical limit \eqref{classical_statistical_limit}, the dynamical KMS symmetry \eqref{dynamical_KMS_symmetry} will yield the same constraints \eqref{quarticKMS} (plus additional constraints among added terms \eqref{added_terms}). However, imposing \eqref{dynamical_KMS_symmetry} under high-temperature limit \eqref{hbar_beta_expansion} will give different constraints. We demonstrate this claiming in appendix \ref{KMS_all}.

\subsection{Fokker-Planck equation from non-Gaussian effective action} \label{derivation_FP}

As explained in \cite{Crossley:2015evo}, inclusion of non-Gaussian terms, such as $L_{\rm SK}^{(4)}$ of \eqref{EFquartic}, would make it inconvenient to perform a Legendre transformation from variable $q_a$ to noise $\theta$, which amounts to saying that it is generically inconvenient to cast non-Gaussian effective action into a stochastic Langevin-type equation\footnote{The inverse problem, i.e., deriving the MSRDJ-type action for a nonlinear version of \eqref{linear_Langevin}-\eqref{noise_Gauss_white}, was recently considered in \cite{Jana:2021niz}. However, it was found that the parameters in the MSRDJ-type action are not in simple correspondence with those in nonlinear Langevin equation.}. In this subsection, we derive Fokker-Planck type equation from the non-Gaussian effective action, for the purpose of future numerical study.

With MSRDJ action for a classical stochastic system, the derivation of Fokker-Planck equation could be found in e.g. the textbook \cite{Kamenev2011}. This mainly relies on the following observation: the partition function based on MSRDJ action is proportional to the probability of finding the system at certain configuration. Here, we will take an alternative approach, say, by considering Hamiltonian formulation of the effective action and making an analogy with quantum mechanics. To illustrate the derivation, we truncate the expansion \eqref{expansion_eta_xi} at leading order so that the effective Lagrangian reads
\begin{align}
L_{\rm SK}^0 = &-M^{\rm T} q^i_a \ddot q^i_r - M^{\rm L} q^z_a \ddot q^z_r  - q^i_a \eta^{\rm T}_0 \dot q^i_r - q^z_a \eta^{\rm L}_0 \dot q^z_r  + \frac{\rm i }{ 2 }\left[ q^i_a \xi^{\rm T}_0 q^i_a + q^z_a \xi^{\rm L}_0 q^z_a \right]- Q B\epsilon_{ij}\dot q^i_r q_a^j \nonumber \\
& + \kappa^{\rm T} \dot q_r^i q_a^i(q_a^j)^2 + \kappa^{\rm L} \dot q_r^z (q_a^z)^3 + \kappa^{\times}_2 \dot q_r^i q_a^i (q_a^z)^2 + \kappa^{\times}_1\dot q_r^z q_a^z (q_a^i)^2 \nonumber\\
&+\frac{\rm i }{ 4!}\left[ \zeta^{\rm T} (q_a^i)^2(q_a^j)^2 + \zeta^{\rm L}(q_a^z)^4+ \zeta^{\times} (q_a^i)^2(q_a^z)^2\right].
\end{align}

In the Lagrangian formulation for the effective theory, $q_r$ satisfies a second order differential equation. Thus, to search for the Hamiltonian formulation, we need to consider particle velocity $\dot q_r$ as another ``coordinate'' \cite{Kamenev2011}, which can be achieved by introducing associated multipliers
\begin{align}
\tilde L_{\rm SK}^0 = &\tilde L_{\rm SK}^0[q_r, \dot q_r; v, \dot v;q_a] \equiv \lambda_a^i (v^i- \dot q_r^i) + \lambda_a^z (v^z- \dot q_r^z) + L_{\rm SK}^0\big|_{\dot q_r^i \to v^i, \dot q_r^z \to v^z} \nonumber \\
=& \lambda_a^i (v^i- \dot q_r^i) + \lambda_a^z (v^z- \dot q_r^z) - M^{\rm T} q^i_a \dot v^i - M^{\rm L} q^z_a \dot v^z  - \eta^{\rm T}_0 q^i_a v^i - \eta^{\rm L}_0 q^z_a v^z - Q B\epsilon_{ij}v^i q_a^j \nonumber \\
& + \frac{\rm i }{ 2 }\left[ q^i_a \xi^{\rm T}_0 q^i_a + q^z_a \xi^{\rm L}_0 q^z_a \right]+ \kappa^{\rm T} v^i q_a^i(q_a^j)^2 + \kappa^{\rm L} v^z (q_a^z)^3 + \kappa^{\times}_2 v^i q_a^i (q_a^z)^2 + \kappa^{\times}_1v^z q_a^z (q_a^i)^2 \nonumber\\
&+\frac{\rm i }{ 4!}\left[ \zeta^{\rm T} (q_a^i)^2(q_a^j)^2 + \zeta^{\rm L}(q_a^z)^4+ \zeta^{\times} (q_a^i)^2(q_a^z)^2\right].
\end{align}
Then, in the path integral based on modified Lagrangian $\tilde L_{\rm SK}^0$, integrating over multipliers $\lambda_a^i, \lambda_a^z$ gives rise to delta functions $\delta(v^i- \dot q_r^i)$ and $\delta (v^z - \dot q_r^z)$ as desired. Now, the conjugate momenta for $q_r$ and $v$ are defined as
\begin{align}
&k^i \equiv -{\rm i} \frac{\partial \tilde L_{\rm SK}^0}{\partial \dot q_r^i} = {\rm i} \lambda_a^i, \qquad \qquad \quad \, \, \, k^z \equiv - {\rm i} \frac{\partial \tilde L_{\rm SK}^0}{\partial \dot q_r^z} = {\rm i} \lambda_a^z, \nonumber \\
&p^i \equiv - {\rm i} \frac{\partial \tilde L_{\rm SK}^0}{\partial \dot v^i} = {\rm i} M^{\rm T} q_a^i, \qquad \qquad p^z \equiv - {\rm i} \frac{\partial \tilde L_{\rm SK}^0}{\partial \dot v^z} = {\rm i} M^{\rm L} q_a^z.
\end{align}
Therefore, in terms of conjugate pairs $(q_r, k)$ and $(v, p)$, the effective action can be rewritten as the anticipated Hamiltonian formulation
\begin{align}
{\rm i} \tilde I^0 \equiv {\rm i} \int dt \tilde L_{\rm SK}^0 = - \int dt \left[ k^i \dot q_r^i + k^z \dot q_r^z + p^i \dot v^i + p^z \dot v^z - H(k, p; q_r, v) \right], \label{I_Hamiltonian}
\end{align}
where the Fokker-Planck Hamiltonian $H$ is
\begin{align}
H =&v^ik^i  + v^z k^z - \tilde \eta^{\rm T}_0 p^i v^i - \tilde \eta^{\rm L}_0 p^z v^z - \tilde Q B \epsilon_{ij}p^j v^i + \frac{ 1 }{ 2 }\left[\tilde \xi^{\rm T}_0 (p^i)^2  + \tilde \xi^{\rm L}_0 (p^z)^2 \right] \nonumber\\
&-\left[\tilde \kappa^{\rm T} (p^i)^2 p^j v^j + \tilde\kappa^{\rm L} (p^z)^3 v^z  + \tilde \kappa^{\times}_2 (p^z)^2 p^i v^i + \tilde \kappa^{\times}_1(p^i)^2 p^z v^z \right]\nonumber\\
& - \frac{1 }{ 4!}\left[\tilde \zeta^{\rm T} (p^i)^2(p^j)^2 + \tilde \zeta^{\rm L} (p^z)^4 + \tilde \zeta^{\times} (p^i)^2(p^z)^2 \right]. \label{FP_H}
\end{align}
For notational simplicity, we introduced tilded coefficients
\begin{align}
&\tilde \eta_0^{\rm T} = \frac{\eta_0^{\rm T}}{M^{\rm T}}, \qquad  \tilde \eta_0^{\rm L} = \frac{\eta_0^{\rm L}}{M^{\rm L}}, \qquad \tilde Q = \frac{Q}{M^{\rm T}}, \qquad \tilde \xi_0^{\rm T} = \frac{\xi_0^{\rm T}}{(M^{\rm T})^2}, \qquad \tilde \xi_0^{\rm L} = \frac{\xi_0^{\rm L}}{(M^{\rm L})^2}, \nonumber \\
&\tilde \kappa^{\rm T} = \frac{\kappa^{\rm T}}{(M^{\rm T})^3}, \qquad \tilde \kappa^{\rm T} = \frac{\kappa^{\rm L}} {(M^{\rm L})^3}, \qquad \tilde \kappa^{\times}_2 = \frac{\kappa^{\times}_2}{M^{\rm T}(M^{\rm L})^2}, \qquad \tilde \kappa^{\times}_1 = \frac{\kappa^{\times}_1}{M^{\rm L}(M^{\rm T})^2}, \nonumber \\
&\tilde \zeta^{\rm T} = \frac{\zeta^{\rm T}}{(M^{\rm T})^4}, \qquad \tilde \zeta^{\rm L} = \frac{\zeta^{\rm L}}{(M^{\rm L})^4}, \qquad \tilde \zeta^{\times} = \frac{\zeta^{\times}}{(M^{\rm T} M^{\rm L})^2}.
\end{align}

In analogy with quantum mechanics, we will ``quantize'' the action \eqref{I_Hamiltonian} by promoting conjugate pairs to operators and impose canonical commutation relations
\begin{align}
[\hat q_r^i, \hat k^j] = \delta^{ij}, \qquad [\hat q_r^z, \hat k^z]= 1, \qquad [\hat v^i, \hat p^j] = \delta^{ij}, \qquad [\hat v^z, \hat p^z] =1,
\end{align}
which, in ``coordinate'' representation $(\hat q_r \to q_r, \hat v \to v)$, could be realized by replacement rule
\begin{align}
\hat k \to -\vec \nabla \equiv \left(-\frac{\partial}{\partial q_r^i}, -\frac{\partial}{\partial q_r^z} \right), \qquad \hat p \to -\vec \nabla_v \equiv \left(-\frac{\partial}{\partial v^i}, -\frac{\partial}{\partial v^z} \right). \label{replace_rule}
\end{align}
Meanwhile, we obtain the Fokker-Planck type equation
\begin{align}
\partial_t \mathcal P \left( q_r, v, t \right) = \hat H\left( \vec\nabla, \vec \nabla_v; q_r, v\right) \mathcal P \left( q_r, v, t \right), \label{FP_equation}
\end{align}
where $\mathcal P \left( q_r, v, t \right)$, analogous to wave function of quantum mechanics, is the probability of finding the system at configuration $(q_r(t), v(t))$ at time $t$. The Fokker-Planck Hamiltonian operator $\hat H$ is obtained from \eqref{FP_H} by making the replacement rule \eqref{replace_rule}. We split $\hat H$ into three parts
\begin{align}
\hat H = \hat H_0 + \hat H_1 + \hat H_2.
\end{align}
Here, the familiar part $\hat H_0$ is
\begin{align}
\hat H_0 \bullet = &-v^i \frac{\partial}{\partial q_r^i} \bullet  - v^z \frac{\partial}{\partial q_r^z} \bullet + \tilde \eta^{\rm T}_0 \frac{\partial}{\partial v^i} (v^i \bullet) + \tilde \eta^{\rm L}_0 \frac{\partial}{\partial v^z} (v^z \bullet) - \tilde Q B \epsilon_{ij} \frac{\partial}{\partial v^j}( v^i \bullet) \nonumber \\
&+ \frac{ 1 }{ 2 }\left( \tilde \xi^{\rm T}_0 \frac{\partial^2}{\partial v^i \partial v^i} + \tilde \xi^{\rm L}_0 \frac{\partial^2}{\partial v^z \partial v^z} \right) \bullet,
\end{align}
which corresponds to a classical stochastic model, such as anisotropic version of \eqref{linear_Langevin}-\eqref{noise_Gauss_white}. The rest pieces $\hat H_1, \hat H_2$ contain higher order velocity derivatives, and could be viewed as non-Gaussian corrections to classical stochastic model. Specifically, $\hat H_1$ represents nonlinear interactions between noise and dynamical variable
\begin{align}
\hat H_1 \bullet = & \tilde \kappa^{\rm T} \left( \frac{\partial}{\partial v^i}\right)^2 \frac{\partial}{\partial v^j}(v^j \bullet) + \tilde\kappa^{\rm L} \left( \frac{\partial} {\partial v^z}\right)^3 (v^z \bullet)  + \tilde \kappa^{\times}_2 \left( \frac{\partial} {\partial v^z}\right)^2 \frac{\partial}{\partial v^i} (v^i \bullet) \nonumber \\
& + \tilde \kappa^{\times}_1 \left( \frac{\partial}{\partial v^i} \right)^2 \frac{\partial} {\partial v^z}(v^z \bullet ).
\end{align}
The last piece $\hat H_2$ corresponds to non-Gaussianity for noise
\begin{align}
\hat H_2 \bullet = - \frac{1 }{ 4!}\left[\tilde \zeta^{\rm T} \left(\frac{\partial}{\partial v^i} \right)^2 \left(\frac{\partial}{\partial v^j}\right)^2 + \tilde \zeta^{\rm L} \left(\frac{\partial}{\partial v^z}\right)^4 + \tilde \zeta^{\times} \left( \frac{\partial} {\partial v^i} \right)^2 \left(\frac{\partial}{\partial v^z}\right)^2 \right] \bullet.
\end{align}

Finally, we briefly discuss the strategy of solving generalized Fokker-Planck equation \eqref{FP_equation}. When non-Gaussian parts $\hat H_1, \hat H_2$ are neglected, \eqref{FP_equation} becomes standard Fokker-Planck equation and has been widely studied in the literature, see e.g. \cite{Risken2011}. Basically, the idea is to first search for the stationary solution and then correct it by introducing time-dependence perturbatively. On top of this, it is possible to explore consequences of non-Gaussian corrections $\hat H_1, \hat H_2$ perturbatively. A detailed study along this direction is beyond the scope of present work and is left as a future project.

\section{Study in a microscopic model} \label{study_holo_model}

In this section we reveal systematic corrections to linear Langevin theory \eqref{linear_Langevin}-\eqref{noise_Gauss_white} from a holographic perspective, hopefully shedding light on understanding properties of strongly coupled quark-gluon plasma. On the one hand, this study will confirm results presented in subsection \ref{construct_eff_action}; on the other hand, we compute various coefficients in \eqref{EFquadratic} and \eqref{EFquartic} as functions of magnetic field, which might be useful for heavy-ion collisions.

\begin{figure}[htbp!]
\centering
\includegraphics[width=0.6\textwidth]{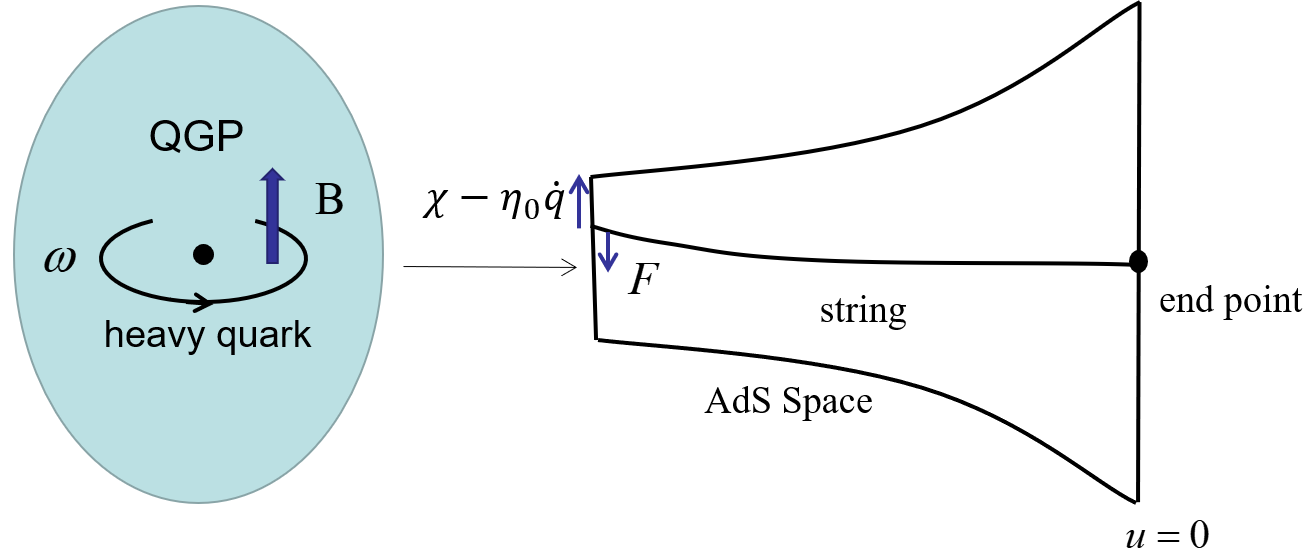}
\caption{Heavy quark in strongly coupled magnetized plasma (left) and its gravity dual (right).} \label{heavyquark}
\end{figure}

Generally, the expectation value of the Wilson loop operator is identified with the partition function of the dual string worldsheet\cite{Maldacena:1998im}. In the large-$N_c$, large-$\lambda$ limit the duality is greatly simplified to
\begin{equation}\label{}
	\langle W(\mathcal{C}) \rangle = e^{i S(\mathcal{C})},
\end{equation}
where $S(\mathcal{C})$ is the Nambu-Goto action, whose boundary condition is that the worldsheet ends on curve $\mathcal{C}$ of the Wilson loop. In our work, we are interested in an unconfined heavy quark, which means we should set the probe D7-brane at the boundary of $\text{AdS}_5$ space. The heavy quark is dual to an open string stretching from the horizon up to the probe D7-brane. Then the Wilson loop turns to the Wilson line, which represents the worldline of the heavy quark.
As in \cite{Herzog:2006gh,Gubser:2006bz,Casalderrey-Solana:2006fio,Mes:2020vgy,Liu:2006he,Liu:2006ug}, we will consider the open string moving in a target space of magnetic AdS$_5$ black brane, which is holographic dual of a heavy quark in strongly coupled magnetized plasma\footnote{In holographic context, dynamics of heavy quark in magnetized plasma was considered in \cite{Kiritsis:2011ha,Finazzo:2016mhm,Li:2016bbh,Dudal:2018rki,Zhang:2018mqt,Kurian:2019nna,Zhu:2019ujc,Arefeva:2020bjk}.  Anisotropic effects on heavy quark dynamics were considered, for e.g., in \cite{Chernicoff:2012iq,Chakrabortty:2013kra,Cheng:2014fza}.  }, see Figure \ref{heavyquark} for illustration.
The friction and noise forces felt by boundary quark correspond to ingoing mode and outgoing one (Hawking mode) in open string's profile \cite{Casalderrey-Solana:2006fio,Son:2009vu,deBoer:2008gu,Giecold:2009cg,CasalderreySolana:2009rm,Atmaja:2010uu}, respectively. Here, adopting the holographic prescription for SK closed time contour \cite{Glorioso:2018mmw}\footnote{In recent years, the holographic SK contour \cite{Glorioso:2018mmw} attracted a lot of attention in various holographic settings \cite{Chakrabarty:2019aeu,Jana:2020vyx,Chakrabarty:2020ohe,Loganayagam:2020eue,
Loganayagam:2020iol,Ghosh:2020lel,Bu:2020jfo,Bu:2021clf,He:2021jna,Bu:2021jlp,Bu:2022esd,He:2022deg}.}, we will extend this picture to nonlinear level (see \cite{Chakrabarty:2019aeu} for the situation without magnetic field) by analyzing dynamics of a Nambu-Goto string in magnetic AdS$_5$ black brane.

The partition function for the bulk theory is
\begin{align}
Z_{\rm bulk} = \int [DX] [D g_{MN}] e^{{\rm i} S_{\rm bulk}[X, \, g_{MN}]},
\end{align}
where $S_{\rm bulk} $ is the total action for the bulk theory, $g_{MN}$ is the metric of target space (magnetic brane in AdS$_5$), and $X$ describes embedding profile of open string in the target space. In probe limit, the target space does not fluctuate. Then, the bulk partition function $Z_{\rm bulk}$ gets reduced into that of an open string in magnetic AdS$_5$ brane:
\begin{align}
Z_{\rm bulk} \simeq  Z_{\rm string}= \int [D X] e^{{\rm i} S[X]}, \label{Z_string}
\end{align}
where $S$ is the total string action. It will be clear that the string embedding profile $X$ is a functional of quark's position $q$, i.e., $X=X[q]$. Thus, the bulk path integral \eqref{Z_string} will be eventually cast into a path integral over the position $q$. We will work in the saddle point approximation:
\begin{align}
Z_{\rm string}= \int [Dq] e^{{\rm i} S[X[q]]},
\end{align}
where $S[X[q]]$ is the on-shell classical string action. The AdS-CFT conjectures that $Z$ of \eqref{Z_Wilson_RG} is equivalent to $Z_{\rm bulk}$. Thus, in the probe limit, the on-shell string action $S[X[q]]$ will be identified with the effective action $I[q]$ for Brownian particle in plasma medium. Therefore, holographic derivation of $I[q]$ boils down to solving the classical equation of motion (EOM) for an open string in magnetic AdS$_5$ brane.

\subsection{Magnetic AdS$_5$ black brane and its field theory dual}

Consider a five dimensional Einstein-Maxwell theory with a negative cosmological constant (the AdS radius is set to unity)
\begin{equation}
S_{\rm EM} = \frac{1 }{2\kappa^2 }\int d^5 x \sqrt{-g}\left(R-\frac{1 }{ 4} F^2 + 12\right). \label{Einstein_Maxwell}
\end{equation}
The equations of motion (EOMs) for bulk theory \eqref{Einstein_Maxwell} read
\begin{align}
& R_{MN} + 4 g_{MN} - \frac{1}{2} F_{PM} F_{~~N}^P + \frac{1}{12} g_{MN} F^2=0, \nonumber \\
& \nabla_M F^{MN} =0. \label{Einstein_Maxwell_eom}
\end{align}
The theory \eqref{Einstein_Maxwell} admits a magnetic brane solution \cite{DHoker:2009mmn}. To utilize the prescription to integrate out the radius coordinate, we work in the ingoing Eddington–Finkelstein coordinates\cite{Glorioso:2018mmw}. Thus the magnetic brane solution ansatz is,
\begin{align}
& ds^2 = r_h^2\left[-\frac{ 2 }{ r_h } du dt -  \frac{U(u)}{u^2} dt^2 + \frac{V(u)}{u^2} \delta_{ij}dx^i dx^j + \frac{W(u)}{u^2} dz^2\right], \quad i,j=1,2, \nonumber \\
& A = B x^1 dx^2 \Longrightarrow F = B dx^1 \wedge dx^2, \label{magnetic_brane_ansatz}
\end{align}
where the AdS boundary is located at $u=0$ and the event horizon is at $u=1$. For simplicity, $r_h$ will be set to unity, and could be restored by dimensional analysis. The Hawking temperature of magnetic brane \eqref{magnetic_brane_ansatz} is
\begin{align}\label{temperature1}
T= - \frac{ U^\prime(u)}{4\pi}\bigg|_{u=1},
\end{align}
where $T$ should be understood as in unit of $r_h$. With the ansatz \eqref{magnetic_brane_ansatz}, bulk EOMs \eqref{Einstein_Maxwell_eom} consist of three dynamical components \eqref{UVWeom} and one constraint \eqref{conseom},
\begin{align}
	0=&U''(u) +U'(u) \left(\frac{V'(u)}{V(u)}+\frac{W'(u)}{2 W(u)}-\frac{5}{u}\right) +U(u) \left(\frac{8}{u^2}-\frac{2 V'(u)}{u V(u)}-\frac{W'(u)}{u W(u)}\right) \nonumber\\
	&-\frac{B^2 u^2}{3 V(u)^2} -\frac{8}{u^2}, \nonumber\\
	0=&V''(u)+ V'(u) \left(\frac{U'(u)}{U(u)}+\frac{W'(u)}{2 W(u)}-\frac{5}{u}\right) +V(u) \left(-\frac{8}{u^2 U(u)}+\frac{8}{u^2} -\frac{2 U'(u)}{u U(u)} \right. \nonumber\\
	&\left.-\frac{W'(u)}{u W(u)}\right)+ \frac{2 B^2 u^2}{3 U(u) V(u)} , \nonumber\\
	0=&W''(u)+ W'(u) \left(\frac{U'(u)}{U(u)}+\frac{V'(u)}{V(u)}-\frac{4}{u}\right) -\frac{W'(u)^2}{2 W(u)}+ W(u) \left(\frac{-\frac{B^2 u^2}{3 V(u)^2} -\frac{8}{u^2}} {U(u)}  \right. \nonumber\\
	&\left. + \frac{8}{u^2}-\frac{2 U'(u)}{u U(u)}-\frac{2 V'(u)}{u V(u)}\right), \label{UVWeom}
\end{align}
\begin{align}
	0=W''(u)+W(u) \left(\frac{2 V''(u)}{V(u)}-\frac{V'(u)^2}{V(u)^2}\right)-\frac{W'(u)^2}{2 W(u)} \label{conseom}.
\end{align}
where, since we have set $r_h=1$ above, $B$ should be understood as $B/r_h^2$.

The bulk metric shall demonstrate asymptotic AdS behavior near $u=0$, which requires
\begin{align}
U(u ) \to 1, \qquad V(u) \to 1, \qquad W(u) \to 1, \qquad {\rm as} ~~u \to 0. \label{AdS_condition}
\end{align}
Indeed, near AdS boundary $u=0$, the metric functions $U,V,W$ are expanded as:
\begin{align}
	&U(u\to 0)= 1+ U_b^1u+ \frac{1}{4}(U_b^1)^2 u^2+ \frac{B^2}{6(V_b^0)^2} u^4\log u  + U_b^4 u^4 + \cdots, \nonumber \\
	&V(u\to 0)= V_b^0 + V_b^0 U_b^1u+ \frac{1}{4}V_b^0(U_b^1)^2 u^2- \frac{B^2}{12 V_b^0} u^4\log u + V_b^4 u^4 + \cdots , \nonumber \\
	&W(u\to 0)= W_b^0 + W_b^0 U_b^1 u + \frac{1}{4}W_b^0 (U_b^1)^2 u^2+ \frac{W_b^0 B^2}{6 (V_b^0)^2} u^4\log u + W_b^4 u^4 \cdots , \label{UVW_asymp}
\end{align}
where we have made use of bulk EOMs \eqref{UVWeom}-\eqref{conseom}. Obviously, the asymptotic boundary conditions \eqref{AdS_condition} only give rise to ``two'' effective requirements! The regularity requirements will yield another three conditions.
Here, as in \cite{Bu:2019qmd} we can utilise the freedom of redefining the radial coordinate $u$ and set $U_b^1=0$. Therefore, the boundary conditions at $u=0$ are
\begin{align} \label{UVW_boundary}
	U'(u=0)=0,\qquad V(u=0)=W(u=0) = 1.
\end{align}
At the horizon $u=1$, we impose regularity condition
\begin{align}
	&U(u=1)=0, \nonumber\\
	&U'(1)V'(1)-8V(1)-2U'(1)V(1)+\frac{2B^2}{3V(1)}=0, \nonumber\\
	&U'(1) W'(1)-2 W(1) U'(1)-8 W(1) -\frac{B^2 W(1)}{3 V(1)^2}=0. \label{UVW_horizon}
\end{align}
Then, near the horizon $u=1$ the metric functions are expanded as
\begin{align}\label{UVW_horizon1}
	&U(u\to 1)= 0+ U^1_h(u-1)+ U_h^2(u-1)^2+\cdots, \nonumber \\
	&V(u\to 1)=V_h^0 + V_h^1(u-1)+ V_h^2 (u-1)^2 +\cdots, \nonumber \\
	&W(u\to 1)= W_h^0+ W_h^1(u-1) + W_h^2(u-1)^2 + \cdots,
\end{align}
where $U_h^1, V_h^0, W_h^0$ are the horizon data and all the rest coefficients are fully fixed in terms of the horizon data. In terms of horizon data, the black hole temperature \eqref{temperature1} is,
\begin{equation}\label{temperature2}
	T=-\frac{  U_h^1 }{ 4\pi }.
\end{equation}

In order to determine the metric functions $U,V,W$, we shall solve bulk EOMs \eqref{UVWeom}-\eqref{conseom} under boundary conditions \eqref{UVW_boundary} and \eqref{UVW_horizon}.

When magnetic field is weak, metric functions $U,V,W$ can be solved analytically \cite{Bu:2019qmd,Basar:2012gh}
\begin{align}
U(u)&=1-u^4+\frac{ 1 }{ 6 }B^2 u^4 \log(u)+\cdots, \nonumber \\
V(u)&=1+\frac{B^2}{48} \text{Li}_2\left(u^4\right)-\frac{B^2(1-u^4) \log(1-u^4)}{12(u^4+3)}+\cdots, \nonumber \\
W(u) &= 1-\frac{B^2}{24} \text{Li}_2\left(u^4\right)+\frac{B^2(1-u^4) \log(1-u^4)}{6(u^4+3)}+\cdots. \label{UVWweakB}
\end{align}
Meanwhile, the black hole temperature \eqref{temperature2} is expanded as,
\begin{equation}\label{temperature3}
		T=\frac{ 1 }{ \pi }\left(1-\frac{B^2 }{ 24 } \right) + \mathcal{O}(B^4).
\end{equation}

For generic value of $B$, the metric functions are known numerically only \cite{DHoker:2009mmn,Ammon:2017ded,Li:2018ufq,Li:2019bgc,Bu:2019qmd,Ammon:2020rvg} . Practically, instead of solving bulk EOMs \eqref{UVWeom}-\eqref{conseom} under boundary conditions \eqref{UVW_boundary} and \eqref{UVW_horizon}, one could take a set of ``convenient'' horizon data and evolve bulk EOMs \eqref{UVWeom}-\eqref{conseom}. More precisely, we will solve bulk EOMs \eqref{UVWeom}-\eqref{conseom} as an initial value problem \eqref{UVW_horizon1} with horizon data taken as (in unit of $r_h=1$)
\begin{align}
U_h^1 = -4, \qquad V_h^0 =1, \qquad W_h^0 =1. \label{initial_value}
\end{align}
Notice that the choice of $U_h^1=-4$ will set $\pi T=r_h(B)$.
Consequently, near AdS boundary $u=0$ the bulk metric behaves as
\begin{align}
ds^2|_{u\to 0} = \frac{1}{u^2} \left[-dt^2 + v(b) \delta_{ij}d\hat x^i d\hat x^j + w(b) d\hat z^2 \right], \label{incorrect_bdy}
\end{align}
which is not the required one \eqref{AdS_condition}. Finally, one obtains correct solution by rescaling of boundary coordinate
\begin{align}
\hat x^i \to x^i /\sqrt{v(b)}, \qquad \hat z \to z/\sqrt{w(b)}.
\end{align}
Here, we use $b$ to denote the magnetic field in the ``incorrect'' boundary metric \eqref{incorrect_bdy}. Then the physical magnetic field $B$ (in unit of $r_h^2$) should be
\begin{align}
	B= \frac{b}{v(b)}.
\end{align}
Finally, we would like to point out that the background solution obtained with initial conditions \eqref{UVW_horizon1} and \eqref{initial_value} does not necessarily satisfy $U_b^1=0$ (cf. \eqref{UVW_asymp}).

The magnetic brane solution \eqref{magnetic_brane_ansatz} is dual to strongly coupled $\mathcal N=4$ SYM plasma exposed to an external magnetic field. In order to add an external magnetic field for boundary theory, we could think of gauging a $U(1)$-subgroup of $R$-symmetry of $\mathcal N=4$ SYM theory \cite{CaronHuot:2006te}. Schematically, the microscopic Lagrangian for the magnetized $\mathcal N=4$ SYM plasma is \cite{Fuini:2015hba}
\begin{align}
S_{\rm th}[\Phi] = S_{\rm SYM, \,\, min.\, coupled} + S_{\rm e.m.} \label{S_th}
\end{align}
where $S_{\rm SYM, min. coupled}$ represents action for $\mathcal N=4$ SYM theory minimally coupled to a U(1) gauge field, and $S_{\rm e.m.}$ is  Maxwell action for U(1) gauge field. Apparently, the thermal bath described by \eqref{S_th} preserves time-reversal symmetry, which plays a crucial role in formulating EFT for a quantum many-body system \cite{Crossley:2015evo}. From bulk perspective, Einstein-Maxwell theory \eqref{Einstein_Maxwell} transparently preserves time-reversal invariance. However, thanks to usage of ingoing EF coordinate system in \eqref{magnetic_brane_ansatz}, the time-reversal symmetry is not simply realized as $t \to -t$, which will become clear in the linearized string solution. The microscopic time-reversal symmetry will be translated into dynamical KMS symmetry \eqref{dynamical_KMS_symmetry} for effective theory for Brownian particle.

\subsection{Dynamics of open string in magnetic brane}

Classical dynamics of open string is described by Nambu-Goto action
\begin{align}
S_{\rm NG}= -\frac{1}{2\pi \alpha^\prime} \int d^2\sigma \sqrt{-h(X)}, \label{S_NG}
\end{align}
where $h$ is determinant of the induced metric $h_{ab}$ on string worldsheet:
\begin{align}
ds^2_{\rm WS}= h_{ab}d\sigma^a \sigma^b = g_{MN}\frac{\partial X^M}{\partial \sigma^a} \frac{\partial X^N}{\partial \sigma^b}. \label{hab_general}
\end{align}
Here, we use $X^M$ to denote embedding of string in the target space \eqref{magnetic_brane_ansatz}. We will take static gauge so that string worldsheet coordinate is $\sigma^a\equiv (\sigma, \tau) = (u,t)$. Then, embedding of the open string is specified by spatial coordinates $X^i(\sigma^a), X^z(\sigma^a)$. In presence of an external Maxwell field, we shall supplement the Nambu-Goto action \eqref{S_NG} by a boundary term:
\begin{align}
S_{\rm bdy} = Q\int d\tau A_M(X) \frac{dX^M}{d\tau} \bigg|_{\rm bdy}. \label{S_bdy}
\end{align}

\begin{figure}[htbp!]
	\centering
	\includegraphics[width=1\textwidth]{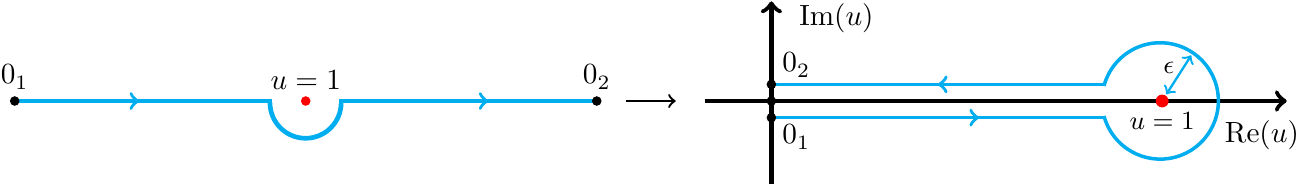}
	\caption{From complexified (analytically continued near horizon) double AdS (left) \cite{Crossley:2015tka} to the holographic SK contour (right) \cite{Glorioso:2018mmw}. The holographic contour infinitely close to the horizon, $\epsilon \to 0$. Indeed, the two horizontal legs overlap with the real axis.} \label{ucontour}
\end{figure}

Imagine a static string with $X^i=X^z=0$, for which the worlsheet spacetime is
\begin{align}
d\bar s_{\rm WS}^2 =- \frac{1}{u^2}\left[ 2 du dt + U(u) dt^2 \right]. \label{WS_AdS}
\end{align}
which has an event horizon identical to that of target space \eqref{magnetic_brane_ansatz}. While such a static string will lose energy into horizon of target space, it will also receive Hawking radiation emitted from the horizon. Resultantly, we will have a fluctuating string around a static configuration. These two modes and their interactions will be translated into effective dynamics of the quark in boundary plasma medium. Following \cite{Glorioso:2018mmw}, we double the background worldsheet spacetime \eqref{WS_AdS}, and analytically continue it around the event horizon $u=1$, so that the radial coordinate $u$ now varies along the contour of Figure \ref{ucontour}.

The specific symmetries postulated for effective action of Brownian particle, say \eqref{Z2_parity}-\eqref{constant_translation}, can be understood from action of open string. Without a nontrivial background, the Nambu-Goto action for open string's fluctuation (still denoted as $X^i, X^z$) preserves the following symmetries independently,
\begin{align}
(1)~~~&X^i \to -X^i, \qquad \,\, \,\,\,\,\, X^z \to -X^z; \nonumber \\
(2)~~~&X^i \to R^{ij} X^j, \qquad \,\,\, X^z \to X^z; \nonumber \\
(3)~~~&X^i \to X^i + c^i, \qquad \, X^z \to X^z + c^z,
\end{align}
which simply translate into \eqref{Z2_parity}-\eqref{constant_translation} at the boundary.

The classical EOM for open string, obtained from \eqref{S_NG}, is highly nonlinear. Thus, we expand \eqref{S_NG} to quartic order in open string's fluctuations $X^i, X^z$:
\begin{align}
S_{\rm NG} = S_{\rm NG}^{(2)} + S_{\rm NG}^{(4)} + \cdots.
\end{align}
The quadratic part of Nambu-Goto action is
\begin{align}
S_{\rm NG}^{(2)} =& -\frac{1 }{2\pi \alpha' } \int dt \int_{0_1}^{0_2} \frac{ du }{ u^2 }~ \Bigg\{V(u)\left[-\partial_{u }X^i \partial_{t }X^i+\frac{ 1}{2 }U(u) (\partial_{u}X^i)^2\right]\nonumber\\
&\qquad\qquad\qquad\qquad\qquad+W(u)\left[-\partial_{u}X^z \partial_{t }X^z+\frac{ 1}{2 }U(u) (\partial_{r }X^z)^2\right]\Bigg\}.  \label{S_NG_quadratic}
\end{align}
The quartic part of Nambu-Goto action is
\begin{align}
S_{\rm NG}^{(4)}=-\frac{ 1 }{ 2\pi \alpha' }\int dt \int_{0_1}^{0_2}  \frac{ du }{ u^2 }\bigg\{&-\frac{ 1 }{ 8 } V(u)^2 \left[2\partial_t X^i - U(u) \partial_u X^i\right]^2 \left(\partial_u X^j\right)^2	\nonumber\\
&-\frac{ 1 }{ 8 } W(u)^2 \left[2\partial_t X^z - U(u) \partial_u X^z\right]^2 \left( \partial_{u}X^z\right)^2\nonumber\\
&-\frac{ 1 }{ 8 }V(u)W(u)\Big[\left(2\partial_t X^i-U(u) \partial_{u }X^i \right)^2 \left(\partial_{u}X^z\right)^2\nonumber\\
& +\left(2\partial_t X^z-U(u) \partial_{u }X^z\right)^2 \left( \partial_{u} X^i \right)^2 \Big] \bigg\}. \label{S_NG_quartic}
\end{align}

Based on truncated action $S_{\rm NG}^{(2)} + S_{\rm NG}^{(4)}$, the EOMs for string fluctuations are
\begin{align}
	&0={\rm{EOMT}}_i=\partial_{u}\left(\frac{U V}{u^2 }\partial_{ u}X^i\right) - \frac{2V} {u^2} \partial_{u} \partial_{t}X^i-\left(\frac{\partial_{u }V} {u^2 }-\frac{2V}{u^3 } \right)\partial_{ t}X^i-f_i[X^i,X^z], \nonumber \\
	&0={\rm{EOML}}_z=\partial_{u}\left(\frac{U W}{u^2 }\partial_{ u}X^z\right) -\frac{2 W} {u^2 } \partial_{u}\partial_{t}X^z - \left(\frac{\partial_{u }W}{u^2 }-\frac{2 W}{u^3 } \right)\partial_{t}X^z-f_z[X^i,X^z],  \label{eom_X}
\end{align}
Here, $f_i, f_z$ are cubic terms in $X^i, X^z$, whose exact forms will not be relevant in subsequent calculations. The string's EOMs \eqref{eom_X} will be solved under doubled AdS boundary conditions
\begin{align}
X^i(t, u= 0_s) = q^i_s, \qquad X^z(t, u=0_s) = q^z_s, \qquad {\rm with}~~ s=1~ {\rm or}~ 2. \label{X_AdS_condition}
\end{align}
The coupled nonlinear system \eqref{eom_X} will be further linearized as
\begin{align}
X^i = \alpha X^i_{(1)} + \alpha^3 X^i_{(3)} + \cdots, \qquad  X^z = \alpha X^z_{(1)} + \alpha^3 X^z_{(3)} + \cdots,
\end{align}
where $\alpha$ is a formal bookkeeping parameter. Accordingly, the boundary conditions \eqref{X_AdS_condition} are implemented as
\begin{align}
&X^i_{(1)}(t, u= 0_s) = q^i_s, \qquad X^z_{(1)}(t, u=0_s) = q^z_s, \qquad {\rm with}~~ s=1~ {\rm or}~ 2, \nonumber \\
&X^i_{(n)}(t, u= 0_s) = 0, \qquad \,\, X^z_{(n)}(t, u=0_s) = 0, \,\,\qquad {\rm with}~~ s=1~ {\rm or}~ 2,~~ {\rm  when }\, \, n\geq 3. \label{X_AdS_condition_perturb}
\end{align}
Then, $X_{(1)}^{i,z}$ satisfy linearized EOMs
\begin{align}
&0=\partial_{u}\left(\frac{U V}{u^2 }\partial_{ u}X^i_{(1)}\right) + {\rm i} \omega\frac{2V} {u^2} \partial_{u} X^i_{(1)}+ {\rm i} \omega\left(\frac{\partial_{u }V} {u^2 }-\frac{2V}{u^3 } \right)X^i_{(1)}, \nonumber \\
&0=\partial_{u}\left(\frac{U W}{u^2 }\partial_{ u}X^z_{(1)}\right) + {\rm i} \omega\frac{2 W} {u^2 } \partial_{u}X^z_{(1)} + {\rm i} \omega\left(\frac{\partial_{u }W}{u^2 }-\frac{2 W}{u^3 } \right)X^z_{(1)},  \label{eom_X_linearized}
\end{align}
where we have turned to frequency domain via Fourier transformation,
\begin{align}
X_{(1)}^{i,z}(u,t) = \int \frac{d\omega}{2\pi} X_{(1)}^{i,z}(u,\omega) e^{-{\rm i} \omega t}.
\end{align}

It turns out that, under AdS boundary conditions \eqref{X_AdS_condition_perturb}, both $S_{\rm NG}^{(2)}$ and $S_{\rm NG}^{(4)}$ are fully determined by linearized fluctuations $X_{(1)}^{i,z}$ \cite{Chakrabarty:2019aeu,Bu:2021jlp}. Indeed, via integration by part, $S_{\rm NG}^{(2)}$ of \eqref{S_NG_quadratic} is reduced into a surface term:
\begin{align}
S_{\rm NG}^{(2)} = -\frac{ 1 }{ 2\pi \alpha^\prime }\int dt \left[\frac{U(u) V(u)}{ 2u^2 }X^i_{(1)}\partial_{ u}X^i_{(1)}+\frac{U(u) W(u)}{ 2u^2 }X^z_{(1)}\partial_{ u}X^z_{(1)}\right]_{0_1}^{0_2} \label{S_NG_quadratic_simple}.
\end{align}
The quadratic order action is simply obtained from \eqref{S_NG_quartic} by replacement rule $X^{i,z} \to X_{(1)}^{i,z}$:
\begin{align}
S_{\rm NG}^{(4)} = -\frac{ 1 }{ 2\pi \alpha' }\int dt \int_{0_1}^{0_2}  \frac{ du }{ u^2 }\bigg\{&-\frac{ 1 }{ 8 } V(u)^2 \left[2\partial_t X^i_{(1)} - U(u) \partial_u X^i_{(1)}\right]^2 \left(\partial_u X^j\right)^2	\nonumber\\
&-\frac{ 1 }{ 8 } W(u)^2 \left[2\partial_t X^z_{(1)} - U(u) \partial_u X^z_{(1)}\right]^2 \left( \partial_{u}X^z_{(1)}\right)^2\nonumber\\
&-\frac{ 1 }{ 8 }V(u)W(u)\Big[\left(2\partial_t X^i_{(1)}-U(u) \partial_{u }X^i_{(1)} \right)^2 \left(\partial_{u}X^z_{(1)}\right)^2\nonumber\\
& +\left(2\partial_t X^z_{(1)}-U(u) \partial_{u }X^z_{(1)}\right)^2 \left( \partial_{u} X^i_{(1)} \right)^2 \Big] \bigg\}.  \label{S_NG_quartic_simple}
\end{align}
Therefore, once linearized profiles $X_{(1)}^{i,z}$ are obtained, evaluating \eqref{S_NG_quadratic_simple} and \eqref{S_NG_quartic_simple} will give effective action for boundary quark.

When $u$ varies along the radial contour of Figure \ref{ucontour}, linearized EOMs \eqref{eom_X_linearized} have been studied in \cite{Chakrabarty:2019aeu,Bu:2021jlp} when $B=0$. The basic idea \cite{Bu:2021jlp} is as follows. First, one cuts the radial contour of Figure \ref{ucontour} at the rightmost point $u= 1 + \epsilon$. It is direct to find out generic solutions when $u$ varies either on upper branch or lower branch of Figure \ref{ucontour}. Then, generic solution on upper branch and generic solution on lower branch will be properly glued at $u= 1 + \epsilon$. The gluing conditions can be derived from the requirement that variational problem of \eqref{S_NG_quadratic} is well-defined at $u= 1 + \epsilon$. Finally, one imposes the AdS boundary conditions \eqref{X_AdS_condition_perturb}. This strategy can be directly applied to solve \eqref{eom_X_linearized} when $B \neq 0$. Below we skip details and present the final solutions.

Under boundary conditions \eqref{X_AdS_condition_perturb}, linearized EOMs \eqref{eom_X_linearized} are solved as
\begin{align}
&X^i_{(1)}(u, \omega) = \mathfrak A_{\rm T}(u, \omega)  q_r^i(\omega) + \mathfrak B_{\rm T}(u, \omega)  q_a^i(\omega), \qquad u \in [0_2, 0_1], \nonumber \\
&X^z_{(1)}(u, \omega) = \mathfrak A_{\rm L}(u, \omega)  q_r^z(\omega) + \mathfrak B_{\rm L}(u, \omega)  q_a^z(\omega), \qquad u \in [0_2, 0_1], \label{X_linearized_solution}
\end{align}
where ($\rm S = T,L$)
\begin{align}
\mathfrak A_{\rm S}(u, \omega) = \frac{\Phi^{\rm ig}_{\rm S}(u, \omega)}{\Phi^{\rm ig(0)}_{\rm S} (\omega)}, \qquad \mathfrak B_{\rm S}(u, \omega) = \frac{1}{2} \coth \frac{\beta\omega}{2} \frac{\Phi^{\rm ig}_{\rm S}(u, \omega)}{\Phi^{\rm ig(0)}_{\rm S} (\omega)} - \frac{e^{2 {\rm i} \omega \chi(u)}}{1-e^{-\beta \omega}} \frac{\Phi^{\rm ig}_{\rm S} (u, -\omega)}{\Phi^{\rm ig(0)}_{\rm S} (-\omega)}, \label{Au_Bu}
\end{align}
where the function $\chi$ is,
\begin{equation}
\chi{(u)}\equiv -\int_{0_2}^{u}\frac{ 1}{ U(u)} \qquad u\in[0_2,0_1].
\end{equation}
Obviously, the task of solving linearized EOMs \eqref{eom_X_linearized} reduces to searching for ingoing solutions $\Phi^{\rm ig}_{\rm T, L}(u, \omega)$. Expressed in the form \eqref{X_linearized_solution}, the linearized solutions demonstrate explicit time-reversal symmetry \cite{Bu:2021clf,Bu:2021jlp}. Importantly, ingoing modes $\Phi^{\rm ig}_{\rm T, L}(u, \omega)$ are regular over the entire radial contour and can be constructed for $u$ varying on either upper branch or lower branch. In the low frequency limit, we have formally constructed $\Phi^{\rm ig}_{\rm T, L}(u, \omega)$,
\begin{align}
\Phi^{\rm ig}_{\rm S}(u,\omega) = \Phi^{\rm ig}_{{\rm S},0}(u) + \omega \Phi^{\rm ig}_{{\rm S},1}(u)  + \omega^2 \Phi^{\rm ig}_{{\rm S},2}(u)  + \cdots, \qquad {\rm S = T,L},   \label{Phi-expand}
\end{align}
where
\begin{align}\label{Phin}
	\Phi_{\rm S,0}^{\rm{ig}}(u)&=1,\nonumber\\
	\Phi_{\rm L,n}^{\rm{ig}}(u)&=-\int_{0}^{u}d\tilde{u}\frac{\tilde{u}^2}{U W }\int_{1}^{\tilde{u}}d\hat{u}~\frac{2{\rm i} W }{\hat{u}^2 }\partial_{\hat{u}}\Phi_{{\rm L},n-1}^{\rm{ig}}+\left(\frac{{\rm i} \partial_{\hat{u} }W}{\hat{u}^2 }-\frac{2{\rm i} W }{\hat{u}^3 }\right)\Phi_{{\rm L},n-1}^{\rm{ig}} , \quad n\geq 1,\\
	\Phi_{\rm T,n}^{\rm{ig}}(u)&=-\int_{0}^{u}d\tilde{u}\frac{\tilde{u}^2}{U V }\int_{1}^{\tilde{u}}d\hat{u}~\frac{2{\rm i} V }{\hat{u}^2 }\partial_{\hat{u}}\Phi_{{\rm T},n-1}^{\rm{ig}}+\left(\frac{{\rm i} \partial_{\hat{u} } V}{\hat{u}^2 }-\frac{2{\rm i} V }{\hat{u}^3 }\right)\Phi_{{\rm T},n-1}^{\rm{ig}} , \qquad n\geq 1.
\end{align}

\subsection{Effective action for Brownian quark: quadratic order}

Quadratic effective action $I^{(2)}$ for boundary quark is related to string action via
\begin{align}
I^{(2)} = S_{\rm NG}^{(2)} + S_{\rm bdy},  \label{I_quadratic_holo}
\end{align}
where $S_{\rm NG}^{(2)}$ is presented in \eqref{S_NG_quadratic_simple}. Near two AdS boundaries, linearized string fluctuations behave as
\begin{align}
X^{i,z}_{(1)}(u \to 0_s, \omega) = q_s^{i,z}(\omega) - {\rm i} \omega q_s^{i,z}(\omega) u + \mathbb{O}^{i,z}_s(\omega) u^3 + \cdots, \qquad s= 1 ~~ {\rm or} ~~ 2, \label{X_i3_bdy}
\end{align}
where the normalizable modes $\mathbb{O}^{i,z}_s(\omega)$ are:
\begin{align}
\mathbb{O}^{i(z)}_2(\omega)= &\frac{\Phi^{\rm ig(3)}_{\rm T(L)}(\omega)}{\Phi^{\rm ig(0)}_{\rm T(L)}(\omega)} q_r^{i(z)}(\omega) + \frac{1}{2}\coth\frac{\beta\omega}{2} \frac{\Phi^{\rm ig(3)}_{\rm T(L)}(\omega)}{\Phi^{\rm ig(0)}_{\rm T(L)}(\omega)} q_a^{i(z)}(\omega) \nonumber \\
& - \frac{1}{1-e^{-\beta\omega}} \frac{\Phi^{\rm ig(3)}_{\rm T(L)}(-\omega)}{\Phi^{\rm ig(0)}_{\rm T(L)}(-\omega)} q_a^{i(z)}(\omega) + \frac{2 {\rm i} \omega^3} {3(1-e^{-\beta\omega})} q_a^{i(z)}(\omega), \nonumber \\
\mathbb{O}^{i(z)}_1(\omega)= &\frac{\Phi^{\rm ig(3)}_{\rm T(L)}(\omega)}{\Phi^{\rm ig(0)}_{\rm T(L)}(\omega)} q_r^{i(z)}(\omega) + \frac{1}{2}\coth\frac{\beta\omega}{2} \frac{\Phi^{\rm ig(3)}_{\rm T(L)}(\omega)}{\Phi^{\rm ig(0)}_{\rm T(L)}(\omega)} q_a^{i(z)}(\omega) \nonumber \\
& - \frac{e^{-\beta\omega}}{1-e^{-\beta\omega}} \frac{\Phi^{\rm ig(3)}_{\rm T(L)} (-\omega)} {\Phi^{\rm ig(0)}_{\rm T(L)}(-\omega)} q_a^{i(z)}(\omega) + \frac{2 {\rm i} \omega^3 e^{-\beta\omega}} {3(1-e^{-\beta\omega})} q_a^{i(z)}(\omega).
\end{align}
where pairing $(i,\rm T)$ or $(z, \rm L)$ is assumed over indices. Here, $\Phi^{\rm ig(0)}$ and $\Phi^{\rm ig(3)}$ are read off from near boundary expansion of ingoing solution
\begin{align}
\Phi^{\rm ig}(u\to 0, \omega) = \Phi^{\rm ig(0)}(\omega) + \cdots + \Phi^{\rm ig(3)}(\omega) u^3+ \cdots.
\end{align}

Immediately, \eqref{I_quadratic_holo} is computed as
\begin{align}
I^{(2)}&= \frac{1}{2\pi \alpha^\prime}\int \frac{d\omega}{2\pi} \left\{ \frac{\rm i}{2}  q_a^i(-\omega)  G_{rr}^{\rm T}(\omega)  q_a^i(\omega) +  q_a^i(-\omega) \left[  M_0 \omega^2 +  G_{ra}^{\rm T}(\omega) \right] q_r^i(\omega) \right\} \nonumber \\
&+\frac{1}{2\pi \alpha^\prime}\int \frac{d\omega}{2\pi} \left\{\frac{\rm i}{2} q_a^3(-\omega)  G_{rr}^{\rm L}(\omega)  q_a^3(\omega) +  q_a^3(-\omega) \left[  M_0 \omega^2 +  G_{ra}^{\rm L}(\omega) \right]  q_r^3(\omega) \right\}\nonumber \\
&-\int \frac{d\omega}{2\pi} Q B i\omega \left[q_r^1(-\omega) q_a^2(\omega) + q_a^1(-\omega) q_r^2(\omega)\right], \label{I_quadratic_holo_final}
\end{align}
where ($\rm S= T, L$)
\begin{align}
& G_{rr}^{\rm S}(\omega)= -{\rm i} \coth\frac{\beta\omega}{2}\left[\frac{3}{2} \frac{\Phi^{\rm ig(3)}_{\rm S}(\omega)} {\Phi^{\rm ig(0)}_{\rm S}(\omega)} - \frac{3}{2} \frac{\Phi^{\rm ig(3)}_{\rm S}(-\omega)} {\Phi^{\rm ig(0)}_{\rm S}(-\omega)} + {\rm i} \omega^3 \right], \nonumber \\
& G_{ra}^{\rm S}(\omega)=3\frac{\Phi^{\rm ig(3)}_{\rm S}(\omega)} {\Phi^{\rm ig(0)}_{\rm S} (\omega)} + {\rm i} \omega^3. \label{Grr_Gra}
\end{align}
In \eqref{I_quadratic_holo_final} the bare quark mass $M_0$ is related to location of probe D7-brane by $M_0=\lim_{u\to \Lambda} 1/u$. The holographic result \eqref{I_quadratic_holo_final} is identical to \eqref{EFquadratic} via the following identification (with $2\pi \alpha^\prime =1$):
\begin{align}
M^{\rm S} \omega^2 + {\rm i} \omega \eta^{\rm S}(\omega) = M_0 \omega^2 + G_{ra}^{\rm S}, \qquad  \xi^{\rm S}=-2 {\rm i} G^{\rm S}_{rr}, \qquad {\rm with ~ S= T~or~L}.
\end{align}
The familiar FDT \eqref{FDT_linear} is equivalent to
\begin{equation}
G_{rr}^{\rm T,L}=\coth{\frac{ \beta \omega}{ 2}} \text{Im} [G_{ra}^{\rm T, L}(\omega)],
\end{equation}
which is automatically satisfied.

While it is straightforward to numerically compute $G_{rr}^{\rm T,L}$ and $G_{ra}^{\rm T,L}$ by scanning over $(\omega, B)$, we will be limited to the leading order results $\eta_0^{\rm T, L}$ (equivalently $\xi_0^{\rm T,L}$), which are related to metric functions by
\begin{align}
&\eta_0^{\rm T}= 1 + \int_1^0 du \left[ \frac{2}{u^3} (W(u)-1) - \frac{W^\prime(u)}{u^2}  \right], \nonumber \\
&\eta_0^{\rm L}= 1 + \int_1^0 du \left[ \frac{2}{u^3} (V(u)-1) - \frac{V^\prime(u)}{u^2}  \right].
\end{align}

\begin{figure}[htbp!]
\centering
\includegraphics[width=0.49\textwidth]{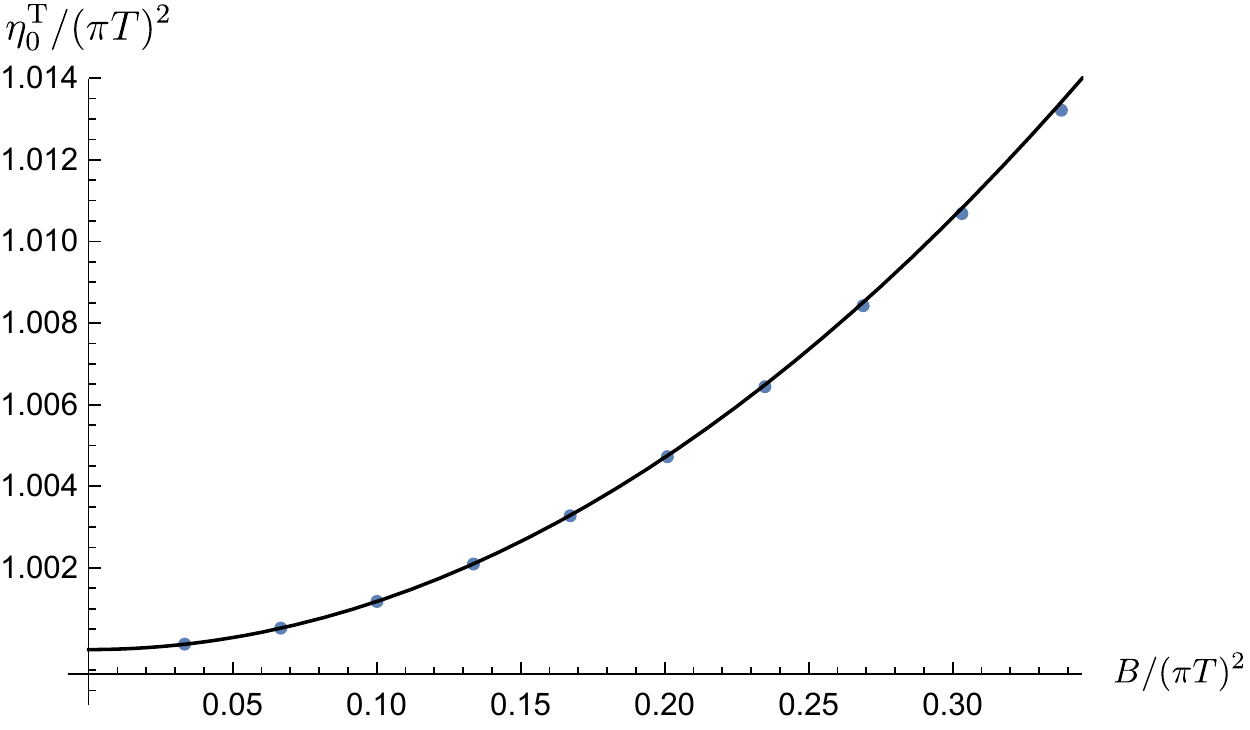}
\includegraphics[width=0.49\textwidth]{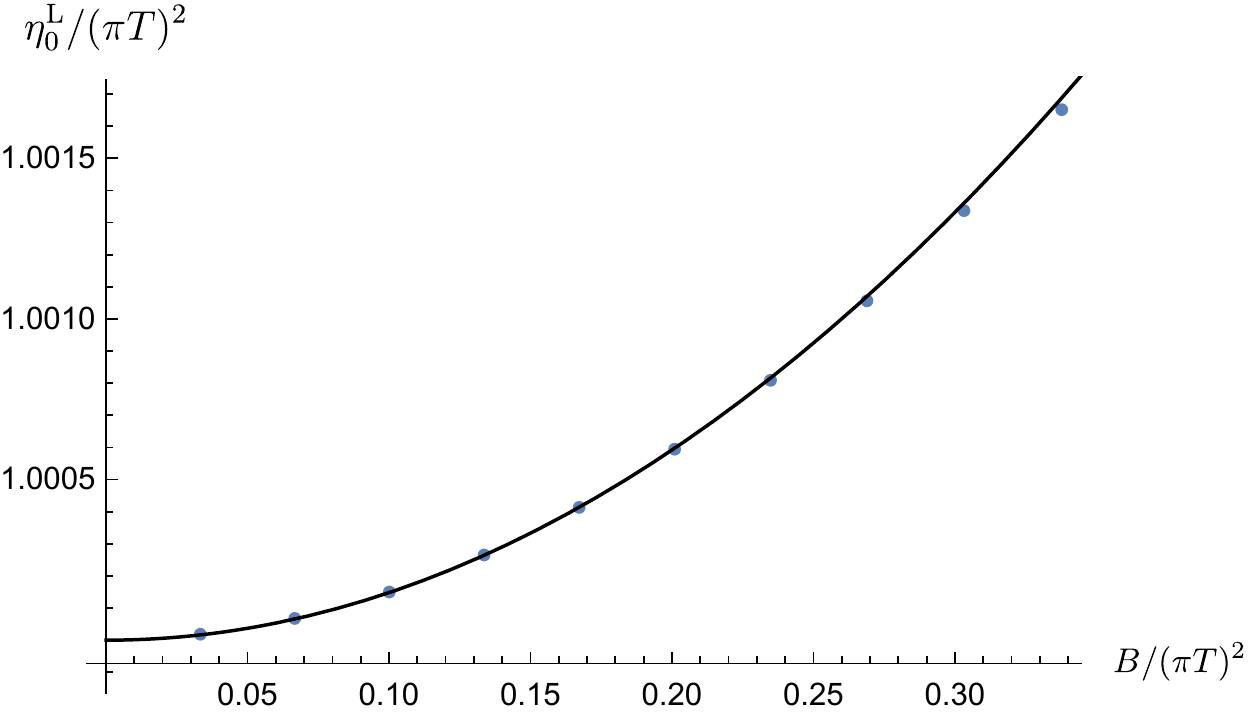}
\caption{Demonstration of perfect agreement between analytical (solid lines) and numerical (dots) results for $\eta_0^{\rm T,L}$ when $B/T^2 \ll 1$.} \label{eta0_weakB}
\end{figure}

When $B/T^2 \ll 1$, metric functions $U, V, W$ are known analytically, see \eqref{UVWweakB}. Thus, we have perturbative expansions for $\eta_0^{\rm T, L}$:
\begin{align}
&\frac{ \eta^{\rm T}_0 }{(\pi T)^2  }=1+\frac{ B^2 }{ 12 (\pi T)^4 }+\frac{  \pi^2 B^2}{ 288 (\pi T)^4} + \cdots, \nonumber\\
&\frac{ \eta^{\rm L}_0 }{(\pi T)^2  }=1+\frac{ B^2 }{ 12 (\pi T)^4 }-\frac{  \pi^2 B^2}{ 144 (\pi T)^4} + \cdots,
\end{align}
which are in perfect agreement with numerical results, as demonstrated in Figure \ref{eta0_weakB}. Here we restore the $r_h$ in $\eta_0^{\rm T,L},B$ and transform the $r_h$ to the temperature of plasma.

\begin{figure}[htbp!]
\centering
\includegraphics[width=0.49\textwidth]{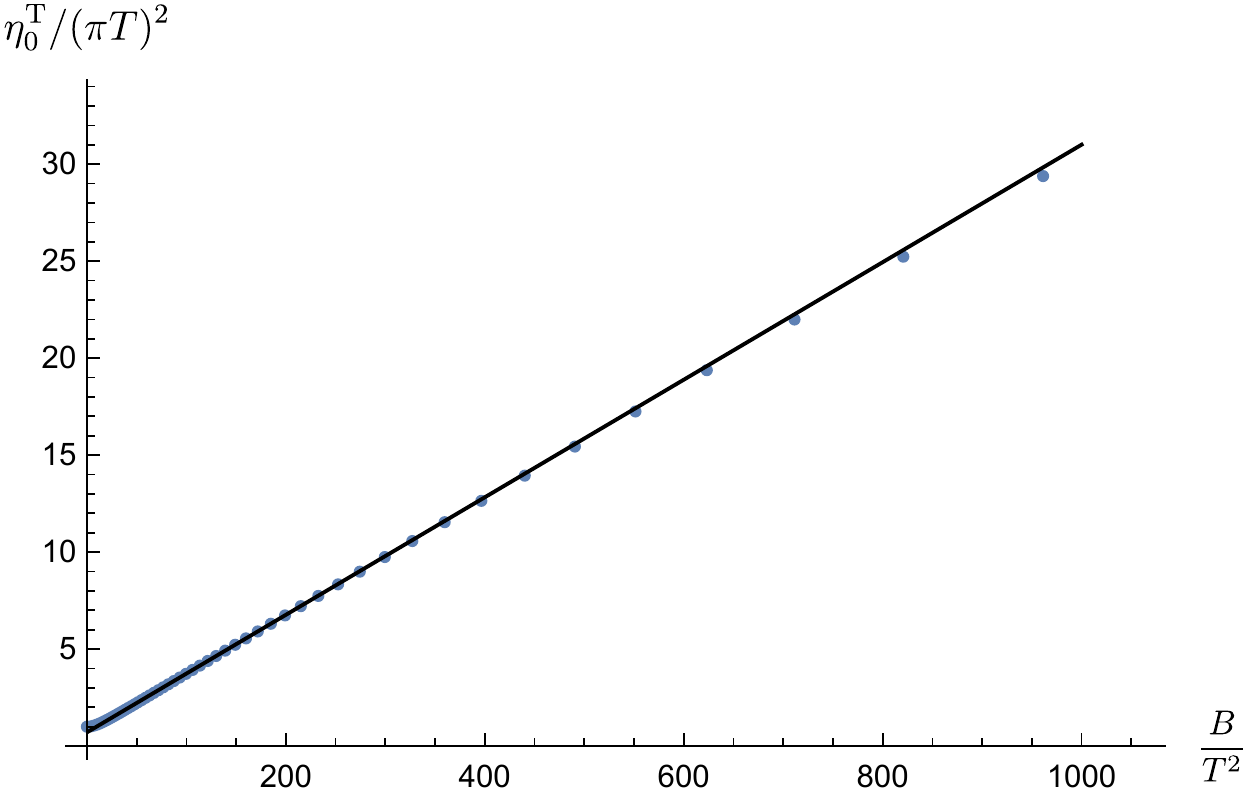}
\includegraphics[width=0.49\textwidth]{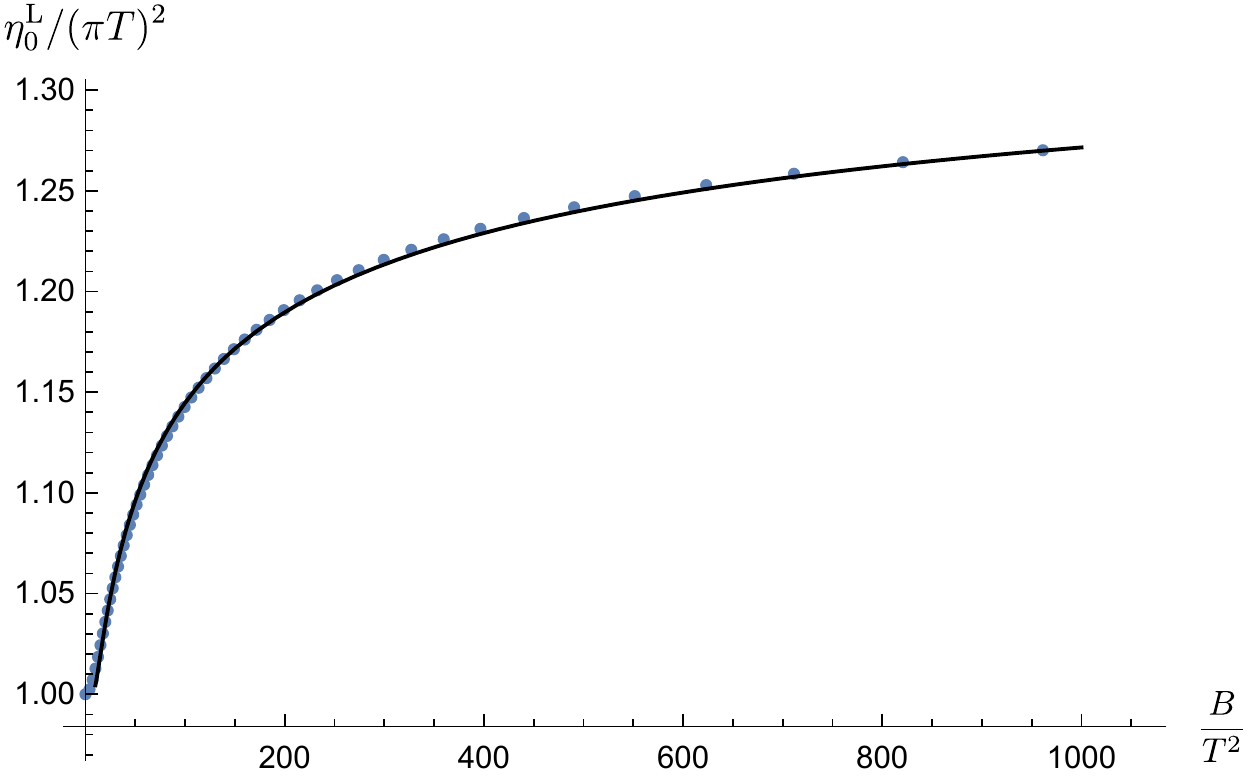}
\caption{Numerical (dots) results for $\eta_0^{\rm T,L}$, which are fitted to simple functions \eqref{eta0_fit} (solid lines) when $B/T^2 \gg 1$.} \label{eta0_genericB}
\end{figure}

\begin{figure}[htbp!]
\centering
\includegraphics[width=0.5\textwidth]{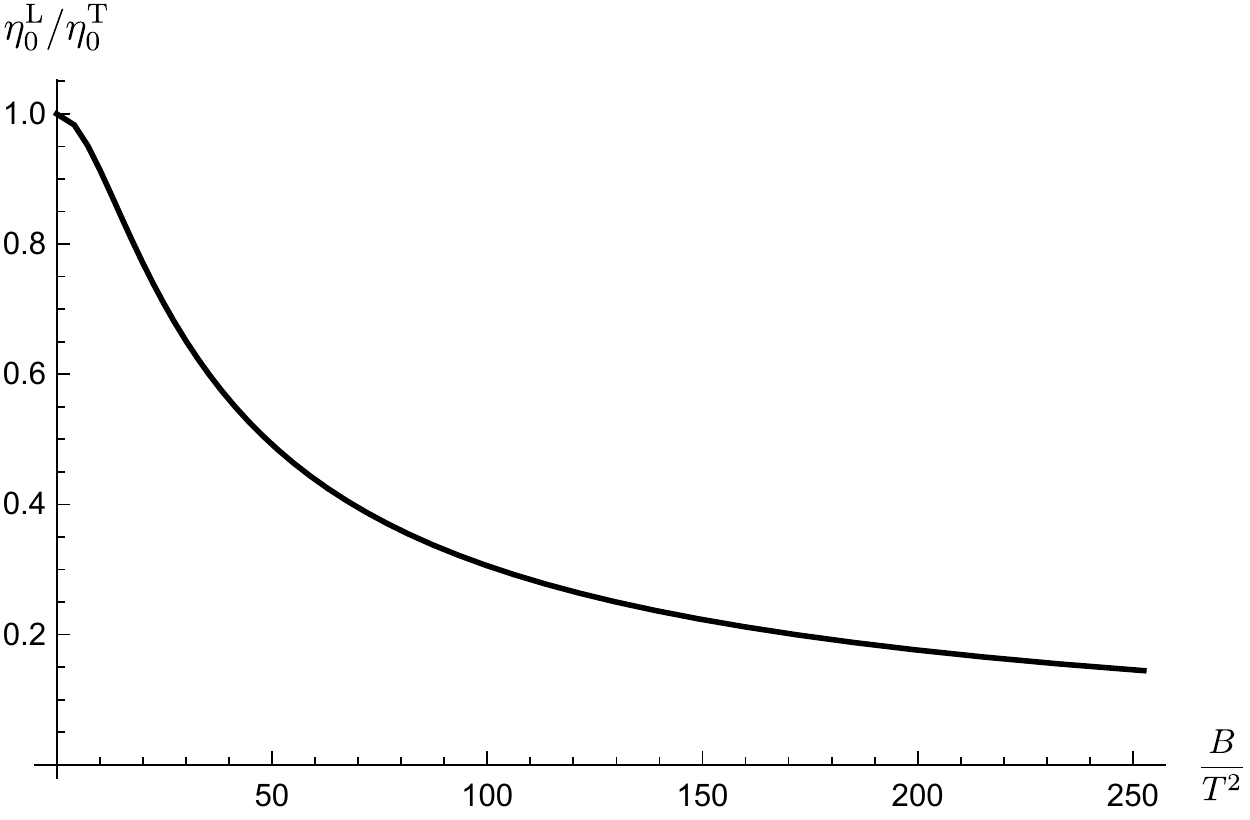}
\caption{Ratio $\eta_0^{\rm L}/\eta_0^{\rm T}$ as a function of $B$, in perfect agreement with that of \cite{Finazzo:2016mhm}.} \label{eta0_ratio}
\end{figure}

For generic value of $B/T^2$, we show numerical results for $\eta_0^{\rm T, L}$ in plots \ref{eta0_genericB} and \ref{eta0_ratio}.
Given that a strong magnetic field is produced in off-center heavy-ion collisions, we examine $\eta_0^{\rm T, L}$ when $B/T^2 \gg 1$, which are well fitted as
\begin{align}
\frac{\eta_0^{\rm T}}{(\pi T)^2} \to 0.707 + 0.0303 \frac{B}{T^2}, \qquad \frac{\eta_0^{\rm L}}{(\pi T)^2} \to 1.386 - 0.524 \frac{\log \left(B/T^2 \right)}{\sqrt{B/T^2}}, \quad {\rm as} ~ B/T^2 \gg 1. \label{eta0_fit}
\end{align}
Apparently, the magnetic field strengths damping effects for both transverse and longitudinal sectors. While the enhancement in transverse sector is linear in $B$ (qualitatively similar to weakly coupled QGP \cite{Fukushima:2015wck}), it seems to saturate for longitudinal mode with an upper bound $\eta_0^{\rm L} \leq 1.386 (\pi T)^2$.

Finally, we compare damping coefficients for transverse and longitudinal modes by plotting the ratio $\eta_0^{\rm L}/\eta_0^{\rm T}$ in Figure \ref{eta0_ratio}, which is in perfect agreement with that of \cite{Finazzo:2016mhm}. Interestingly, the ratio $\eta_0^{\rm L}/\eta_0^{\rm T}$ shows a reasonably slow decreasing as $B$ becomes large.

\subsection{Effective action for Brownian quark: quartic order}

The quartic effective action $I^{(4)}$ for Brownian quark is related to string action via
\begin{align}
I^{(4)} = S_{\rm NG}^{\rm (4)} \big|_{X^{i,z} \to X^{i,z}_{(1)}},
\end{align}
with $S_{\rm NG}^{\rm (4)}$ presented in \eqref{S_NG_quadratic}. In frequency domain, quartic action becomes convolution
\begin{align}
I^{(4)} = \frac{1}{2\pi \alpha^\prime} \int \frac{d\omega_1 d\omega_2 d\omega_3} {(2\pi)^3} \delta(\omega_1+ \omega_2+ \omega_3+ \omega_4) \sum_{m=1}^{4} \mathcal I_m(\omega_1, \omega_2, \omega_3, \omega_4),  \label{I_quartic_holo}
\end{align}
where
\begin{align}
\mathcal{I}_1=\int_{0_1}^{0_2} & \frac{ du }{ u^2 }\frac{ V(u)^2 }{ 8 } \left[2{\rm i} \omega_1 X_{(1)}^i(u,\omega_1)+U(u) \partial_{u }X_{(1)}^i(u,\omega_1)\right]\nonumber\\
\times &\left[2{\rm i}\omega_2 X_{(1)}^i(u,\omega_2)+U(u) \partial_{u } X_{(1)}^i (u,\omega_2) \right] \partial_{u}X_{(1)}^j(u,\omega_3)\partial_{u} X_{(1)}^j(u,\omega_4), \nonumber\\
\mathcal{I}_2=\int_{0_1}^{0_2} & \frac{ du }{ u^2 }\frac{ W(u)^2 }{ 8 } \left[2{\rm i} \omega_1 X_{(1)}^z(u,\omega_1) + U(u) \partial_{u } X_{(1)}^z (u,\omega_1)\right] \nonumber\\
\times &\left[2{\rm i}\omega_2 X_{(1)}^z(u,\omega_2)+ U(u) \partial_{u }X_{(1)}^z (u,\omega_2) \right] \partial_{u}X_{(1)}^z (u,\omega_3) \partial_{u} X_{(1)}^z (u,\omega_4), \nonumber \\
\mathcal{I}_3=\int_{0_1}^{0_2}& \frac{ du }{ u^2 }\frac{ W(u)V(u) }{ 8 } \left[2{\rm i} \omega_1 X_{(1)}^i(u,\omega_1)+U(u) \partial_{u }X_{(1)}^i(u,\omega_1)\right] \nonumber\\
\times &\left[2{\rm i}\omega_2 X_{(1)}^i(u,\omega_2)+U(u) \partial_{u } X_{(1)}^i (u,\omega_2)\right] \partial_{u}X_{(1)}^z(u,\omega_3)\partial_{u}X_{(1)}^z (u,\omega_4), \nonumber \\
\mathcal{I}_4=\int_{0_1}^{0_2} & \frac{ du }{ u^2 }\frac{ W(u)V(u) }{ 8 } \left[2{\rm i} \omega_1 X_{(1)}^z(u,\omega_1)+U(u) \partial_{u }X_{(1)}^z(u,\omega_1)\right] \nonumber\\
\times & \left[2{\rm i}\omega_2 X_{(1)}^z(u,\omega_2)+U(u) \partial_{u } X_{(1)}^z (u,\omega_2) \right] \partial_{u}X_{(1)}^i(u,\omega_3)\partial_{u}X_{(1)}^i(u,\omega_4). \label{I_integral1-4}
\end{align}
In contrast to computation of \eqref{I_quadratic_holo}, quartic action \eqref{I_quartic_holo} inevitably involves contour integrals \eqref{I_integral1-4}, which are generically hard to compute. We proceed by examining singular behavior for integrands of \eqref{I_integral1-4} when $u$ is in the region enclosed by the radial contour. Recall that $X_{(1)}^{i,z}$ is linear superposition of ingoing mode and Hawking mode \eqref{X_linearized_solution}: while the former is regular when $u$ is inside the contour, the latter shows logarithmic singularity near horizon due to the oscillating factor $e^{2i\omega \chi(u)}$ (see \eqref{Au_Bu})
\begin{align}
e^{2{\rm i}\omega \chi(u)} = (u-1)^{{\rm i}\beta \omega/(2\pi)} g(u, \omega),
\end{align}
where $g(u,\omega)$ is a regular function of $u$. This type of singularity raises potential subtlety \cite{Glorioso:2018mmw} regarding the order of taking $\omega \to 0$ limit and taking the limit $\epsilon \to 0$  (cf. Figure \ref{ucontour} for $\epsilon$)
\begin{equation}
\lim_{\omega \to 0}\lim_{\epsilon\to 0}e^{2{\rm i}\omega \chi(u)}\neq \lim_{\epsilon\to 0}\lim_{\omega \to 0}e^{2{\rm i}\omega \chi(u)},  \qquad \text{as}\quad u\to 1+\epsilon. \label{non-commute-issue}
\end{equation}
However, for the purpose of calculating \eqref{I_quartic_holo} up to $\mathcal{O}(\omega^1)$, it turns out that non-commutativity issue of \eqref{non-commute-issue} is accidentally washed away, as demonstrated in appendix \ref{non-commutativity}. Thus, it becomes valid to proceed as follows: expand integrands of \eqref{I_integral1-4} in low frequency limit, and then evaluate radial integrals at each order in $\omega$, and finally take the limit $\epsilon \to 0$.

To facilitate discussion of derivative expansion for \eqref{I_quartic_holo}, we introduce compact notations for each piece in \eqref{I_integral1-4}
\begin{align}
& 2{\rm i} \omega X_{(1)}^i + U(u)\partial_{u } X_{(1)}^i \equiv \hat{\mathcal A}_{\rm T} q^i_r + \hat{\mathcal B}_{\rm T} q^i_a,\qquad \qquad  \partial_{u }X_{(1)}^i \equiv \tilde{\mathcal A}_{\rm T} q^i_r + \tilde{\mathcal B}_{\rm T} q^i_a,\nonumber\\
& 2{\rm i}\omega X_{(1)}^z + U(u) \partial_{u} X_{(1)}^z \equiv \hat{\mathcal A}_{\rm L} q^z_r + \hat{\mathcal B}_{\rm L} q^z_a, \qquad \qquad \partial_{u }X_{(1)}^z \equiv \tilde{\mathcal A}_{\rm L} q^z_r + \tilde{\mathcal B}_{\rm L} q^z_a, \label{hattildeAB}
\end{align}
where ($\rm S = T, L$),
\begin{align}
\hat{\mathcal A}_{\rm S} = 2{\rm i}\omega \mathfrak A_{\rm S} + U(u)\partial_{ u} \mathfrak A_{\rm S}, \quad \hat{\mathcal B}_{\rm S} = 2{\rm i}\omega \mathfrak B_{\rm S} + U(u)\partial_{ u} \mathfrak B_{\rm S}, \quad \tilde{\mathcal A}_{\rm S} = \partial_{ u} \mathfrak A_{\rm S}, \quad \tilde{\mathcal B}_{\rm S}=\partial_{ u} \mathfrak B_{\rm S}. \label{ABhattilde}
\end{align}
In terms of $\hat{\mathcal A}_{\rm T,L}, \hat{\mathcal B}_{\rm T,L}, \tilde{\mathcal A}_{\rm T,L}, \tilde{\mathcal B}_{\rm T,L}$, contour integrals in \eqref{I_integral1-4} become
\begin{align}
\mathcal I_1&=\int_{0_1}^{0_2}du\frac{ V(u)^2}{8 u^2 }\biggl\{ \hat{\mathcal B}_{\rm T} (\omega_1) \hat{\mathcal B}_{\rm T} (\omega_2) \tilde{\mathcal B}_{\rm T} (\omega_3) \tilde{\mathcal B}_{\rm T} (\omega_4) q_{a}^i (\omega_1) q_{a}^i (\omega_2) q_{a}^j (\omega_3) q_{a}^j (\omega_4) \nonumber\\
&\qquad\qquad\qquad \quad +2\hat{\mathcal A}_{\rm T} (\omega_1) \hat{\mathcal B}_{\rm T} (\omega_2) \tilde{\mathcal B}_{\rm T} (\omega_3) \tilde{\mathcal B}_{\rm T} (\omega_4) q_{r}^i (\omega_1) q_{a}^i (\omega_2)q_{a}^j(\omega_3) q_{a}^j(\omega_4)\nonumber\\
&\qquad\qquad\qquad \quad +2\hat{\mathcal B}_{\rm T} (\omega_1) \hat{\mathcal B}_{\rm T} (\omega_2) \tilde{\mathcal A}_{\rm T} (\omega_3) \tilde{\mathcal B}_{\rm T} (\omega_4) q_{a}^j (\omega_1) q_{a}^j (\omega_2) q_{r}^i(\omega_3)q_{a}^i(\omega_4) \biggr\} \nonumber,\\
\mathcal I_2&=\int_{0_1}^{0_2}du\frac{ W(u)^2}{8 u^2 }\biggl\{\hat{\mathcal B}_{\rm L} (\omega_1) \hat{\mathcal B}_{\rm L} (\omega_2) \tilde{\mathcal B}_{\rm L} (\omega_3) \tilde{\mathcal B}_{\rm L} (\omega_4) q_{a}^3 (\omega_1) q_{a}^3 (\omega_2) q_{a}^3 (\omega_3) q_{a}^3(\omega_4)\nonumber\\
&\qquad\qquad\qquad \quad + 2\hat{\mathcal A}_{\rm L} (\omega_1) \hat{\mathcal B}_{\rm L} (\omega_2) \tilde{\mathcal B}_{\rm L} (\omega_3) \tilde{\mathcal B}_{\rm L} (\omega_4) q_{r}^3 (\omega_1) q_{a}^3 (\omega_2) q_{a}^3(\omega_3) q_{a}^3 (\omega_4) \nonumber\\
&\qquad\qquad\qquad \quad + 2 \hat{\mathcal B}_{\rm L} (\omega_1) \hat{\mathcal B}_{\rm L} (\omega_2) \tilde{\mathcal A}_{\rm L} (\omega_3) \tilde{\mathcal B}_{\rm L} (\omega_4) q_{a}^3 (\omega_1) q_{a}^3 (\omega_2) q_{r}^3 (\omega_3) q_{a}^3 (\omega_4) \biggr\} \nonumber,\\
\mathcal I_3&=\int_{0_1}^{0_2}du\frac{ W(u)V(u)}{8 u^2 }\biggl\{ \hat{\mathcal B}_{\rm T} (\omega_1) \hat{\mathcal B}_{\rm T} (\omega_2) \tilde{\mathcal B}_{\rm L} (\omega_3) \tilde{\mathcal B}_{\rm L} (\omega_4) q_{a}^i (\omega_1) q_{a}^i (\omega_2) q_{a}^3 (\omega_3) q_{a}^3(\omega_4) \nonumber\\
&\qquad\qquad\qquad\qquad \quad +2\hat{\mathcal A}_{\rm T} (\omega_1) \hat{\mathcal B}_{\rm T} (\omega_2) \tilde{\mathcal B}_{\rm L} (\omega_3) \tilde{\mathcal B}_{\rm L} (\omega_4) q_{r}^i(\omega_1) q_{a}^i (\omega_2) q_{a}^3(\omega_3) q_{a}^3(\omega_4) \nonumber\\
&\qquad\qquad\qquad \qquad \quad + 2\hat{\mathcal B}_{\rm T} (\omega_1) \hat{\mathcal B}_{\rm T} (\omega_2) \tilde{\mathcal A}_{\rm L} (\omega_3) \tilde{\mathcal B}_{\rm L} (\omega_4) q_{a}^i(\omega_1) q_{a}^i (\omega_2) q_{r}^3(\omega_3) q_{a}^3(\omega_4) \biggr\} \nonumber,\\
\mathcal I_4&=\int_{0_1}^{0_2}du\frac{ W(u)V(u)}{8 u^2 }\biggl\{\hat{\mathcal B}_{\rm L} (\omega_1) \hat{\mathcal B}_{\rm L} (\omega_2) \tilde{\mathcal B}_{\rm T} (\omega_3) \tilde{\mathcal B}_{\rm T} (\omega_4) q_{a}^3(\omega_1) q_{a}^3(\omega_2) q_{a}^i(\omega_3) q_{a}^i(\omega_4) \nonumber\\
&\qquad\qquad\qquad \qquad \quad +2\hat{\mathcal A}_{\rm L} (\omega_1) \hat{\mathcal B}_{\rm L} (\omega_2) \tilde{\mathcal B}_{\rm T} (\omega_3) \tilde{\mathcal B}_{\rm T} (\omega_4) q_{r}^3(\omega_1) q_{a}^3(\omega_2) q_{a}^i(\omega_3)q_{a}^i(\omega_4)\nonumber\\
&\qquad\qquad\qquad \qquad \quad +2\hat{\mathcal B}_{\rm L} (\omega_1) \hat{\mathcal B}_{\rm L} (\omega_2) \tilde{\mathcal A}_{\rm T} (\omega_3) \tilde{\mathcal B}_{\rm T} (\omega_4) q_{a}^3(\omega_1) q_{a}^3(\omega_2) q_{r}^i(\omega_3) q_{a}^i(\omega_4) \biggr\} \label{Intfunc},
\end{align}
where we omitted terms that are explicitly beyond $\mathcal{O}(\omega^1)$.

To leading order in $\omega$, the coefficients in \eqref{X_linearized_solution} are expanded as ($\rm S=T,L$)
\begin{align}\label{ABlowfrequency}
\mathfrak A_{S}= 1+\mathcal{O}(\omega), \qquad\qquad
\mathfrak B_{S}=-\frac{ 1 }{ 2 }+\frac{2}{\beta} \frac{\Phi^{\rm ig}_{{\rm S},1} (u) }{\Phi^{\rm ig}_{{\rm S},0}(u) }-\frac{2\rm i}{\beta} \chi(u)+\mathcal{O}(\omega).
\end{align}
where $\Phi^{\rm ig}_{{\rm S},0} (u)$ and $\Phi^{\rm ig}_{{\rm S},1} (u)$ are introduced in low frequency expansion of ingoing solution $\Phi^{\rm ig}(u,\omega)$, see \eqref{Phi-expand}. So, in low frequency limit, \eqref{ABhattilde} scale as
\begin{equation}
\tilde{\mathcal A}_{\rm T,L},~ \hat{\mathcal A}_{\rm T,L}\sim \mathcal{O}(\omega^1), \qquad \tilde{\mathcal B}_{\rm T,L}, ~ \hat{\mathcal B}_{\rm T,L}\sim \mathcal{O}(\omega^0).  \label{ABhattilde_scaling}
\end{equation}

To extract $\mathcal{O}(\omega^1)$ part of $I^{(4)}$, in \eqref{Intfunc} it is sufficient to retain $\hat{\mathcal B}\hat{\mathcal B}\tilde{\mathcal B}\tilde{\mathcal B}$ type terms to $\mathcal{O}(\omega^0)$ while retain $\hat{\mathcal A}\hat{\mathcal B}\tilde{\mathcal B}\tilde{\mathcal B}$ type terms to $\mathcal{O}(\omega^1)$. This means that we only need lowest order terms of  $\tilde{\mathcal A}_{T,L}$, $\hat{\mathcal A}_{T,L}$, $\tilde{\mathcal B}_{T,L}$, $\hat{\mathcal B}_{T,L}$, all of which are regular. Therefore, expressed in the form \eqref{Intfunc}, it is transparent that the contour integrals can be computed by residue theorem. Eventually, the result \eqref{EFquartic} is recovered with various coefficients given as
\begin{align}
&  \kappa^{\rm T} = \frac{B^2-24 (V_h^0)^2}{32 \pi ^3 \alpha^\prime }, \qquad \qquad \qquad \qquad \zeta^{\rm T} =\frac{-12 \beta\left[B^2-24 (V_h^0)^2\right]}{32 \pi ^3 \alpha^\prime }, \nonumber \\
& \kappa^{\rm L} = -\frac{3 (W_h^0)^2 \left[B^2+8 (V_h^0)^2\right]}{32 \pi ^3 \alpha^\prime  (V_h^0)^2}, \qquad \quad\,\,\,  \zeta^{\rm L} = \frac{36\beta (W_h^0)^2 \left[B^2+8 (V_h^0)^2\right]}{32 \pi ^3 \alpha^\prime  (V_h^0)^2}, \nonumber \\
&  \kappa^\times_{1,2} = -\frac{W_h^0 \left[B^2+24 (V_h^0)^2\right]}{32 \pi ^3 \alpha^\prime  V_h^0}, \qquad \quad~\,\,\,\, \zeta^\times =  \frac{24\beta W_h^0 \left[B^2+24 (V_h^0)^2\right]}{32 \pi ^3 \alpha^\prime  V_h^0},  \label{kappa_horizon_data}
\end{align}
where $ V_h^0, W_h^0$ are horizon data, cf.~\eqref{UVW_horizon1}, and we have transformed $U_h^1$ to the inverse temperature $\beta$ by \eqref{temperature2}. Interestingly, the KMS conditions \eqref{quarticKMS} are perfectly satisfied even without knowledge of exact solution for metric functions.

\begin{figure}[htbp!]
	\centering
	\includegraphics[width=0.48\textwidth]{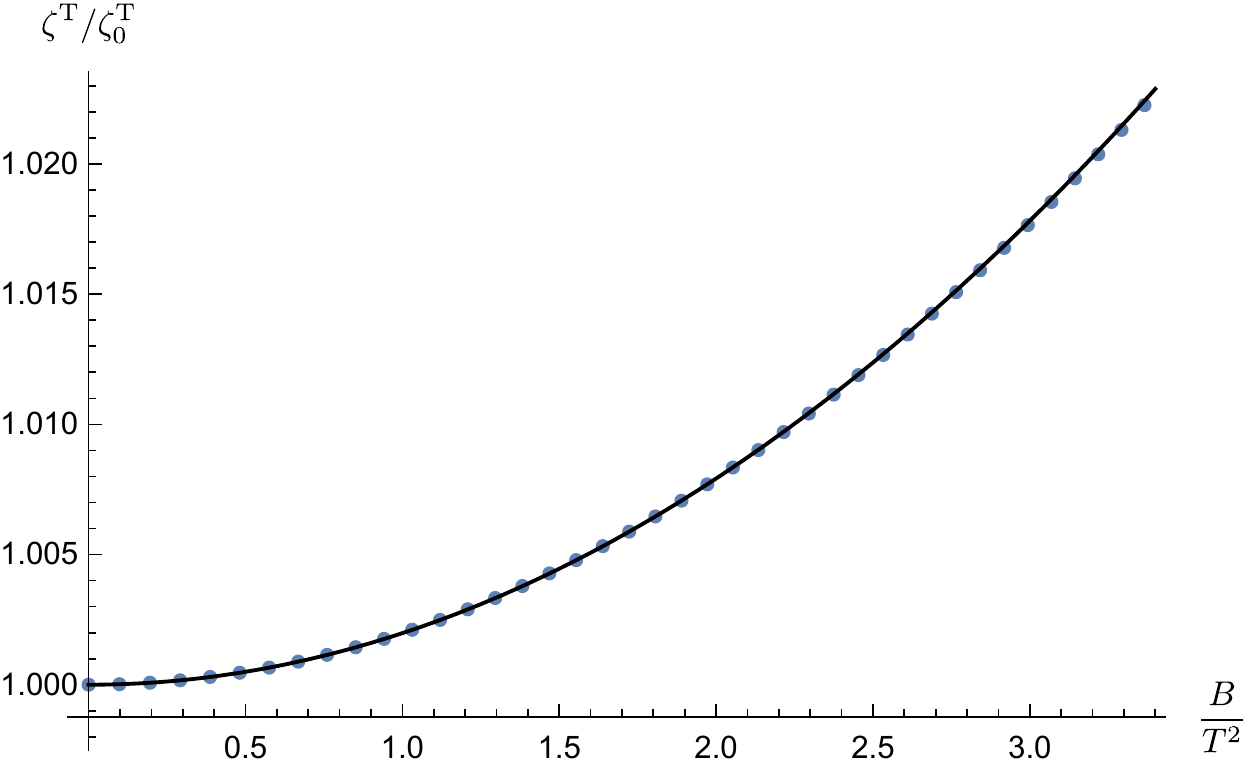}
	\includegraphics[width=0.48\textwidth]{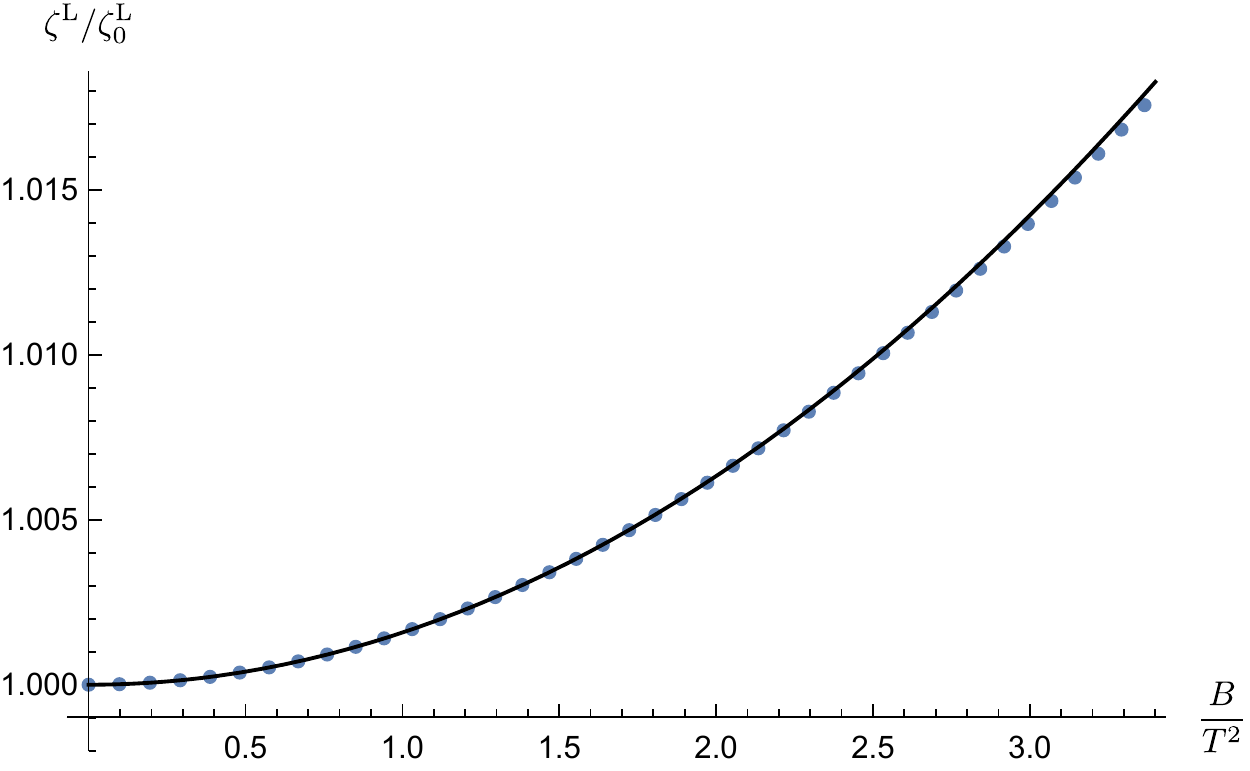}
	\includegraphics[width=0.48\textwidth]{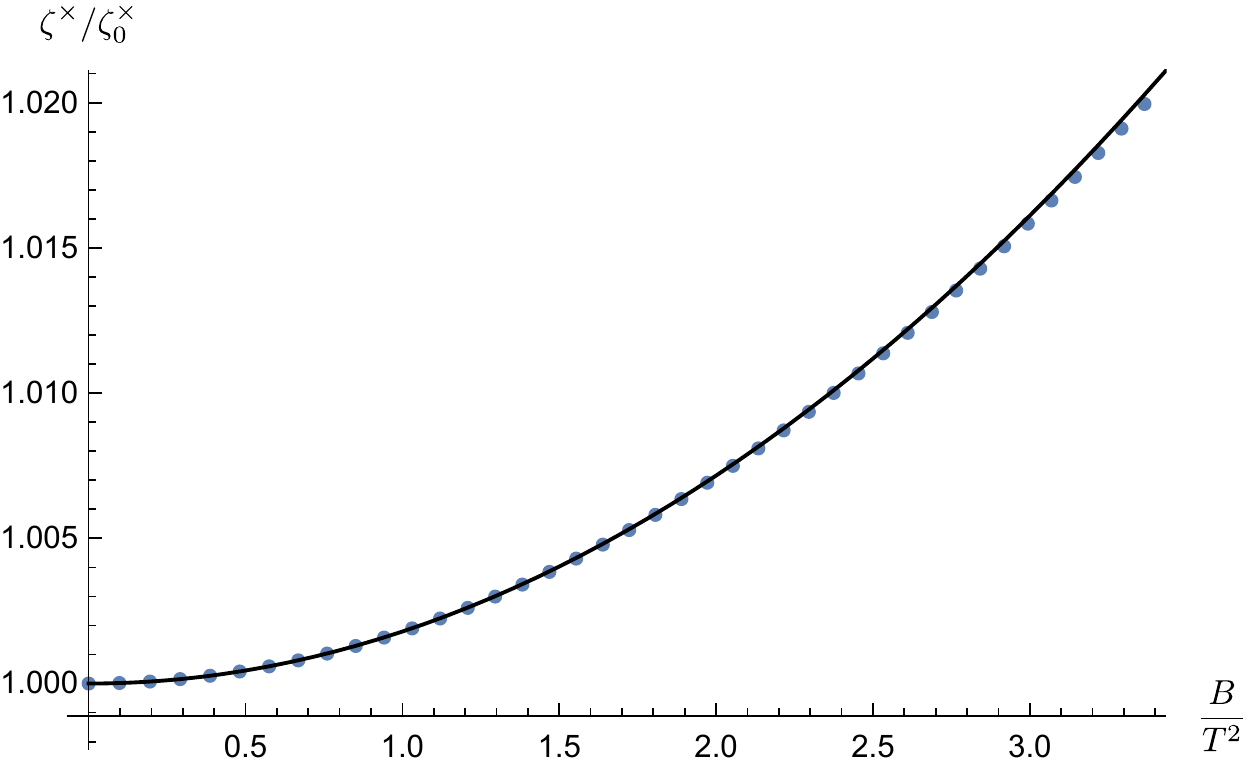}
	\caption{Demonstration of perfect agreement between analytical (solid lines) and numerical (dots) results for $\zeta^{\rm{T,L},\times}/\zeta^{\rm{T,L},\times}_{0},$ when $B/T^2 \ll 1$.} \label{figure_Grrrrweak}
\end{figure}

In weak field limit $B/T^2 \ll 1$, we analytically compute all coefficients in \eqref{kappa_horizon_data}
\begin{align}
&\frac{\zeta^{\rm T}}{\zeta^{\rm T}_{0}} =  1+ \frac{18+ \pi^2} {144} \frac{B^2}{\pi^4 T^4} + \cdots, \nonumber \\
&\frac{\zeta^{\rm L}}{\zeta^{\rm L}_{0}} = 1+ \frac{21- \pi^2} {72} \frac{B^2}{\pi^4 T^4} + \cdots, \nonumber \\
&\frac{\zeta^\times}{\zeta^\times_{0}} = 1+ \frac{60- \pi^2} {288} \frac{B^2}{\pi^4 T^4} + \cdots, \label{zetas}
\end{align}
where $\zeta^{\rm T}_{0},\zeta^{\rm L}_{0},\zeta^{\times}_{0}$ are values of $\zeta^{\rm T},\zeta^{\rm L},\zeta^{\times}$ when $B=0$,
\begin{equation}
	\zeta^{\rm T}_{0}=\frac{9 \pi }{\alpha^\prime  \beta ^5},\qquad
	\zeta^{\rm L}_{0}=\frac{9 \pi }{\alpha^\prime  \beta ^5},\qquad
	\zeta^{\times}_{0}=\frac{18 \pi }{\alpha^\prime  \beta ^5}.\label{zeta0s}
\end{equation}
Here, we have restored the $r_h$ in $B$, $\zeta^{\rm{T,L}\times}$, $\zeta^{\rm{T,L}\times}_0$ and transformed it to temperature in \eqref{zetas} and \eqref{zeta0s}. When $B=0$, our result \eqref{zeta0s} is in agreement with \cite{Chakrabarty:2019aeu}, up to an overall sign. However, we are confident that our result \eqref{zeta0s} is more reasonable once the condition \eqref{imaginary_positive_quartic} is concerned. As shown in Figure \ref{figure_Grrrrweak}, our analytical result \eqref{zetas} is perfectly consistent with numerical study.

\begin{figure}[htbp!]
	\centering
	\includegraphics[width=0.48\textwidth]{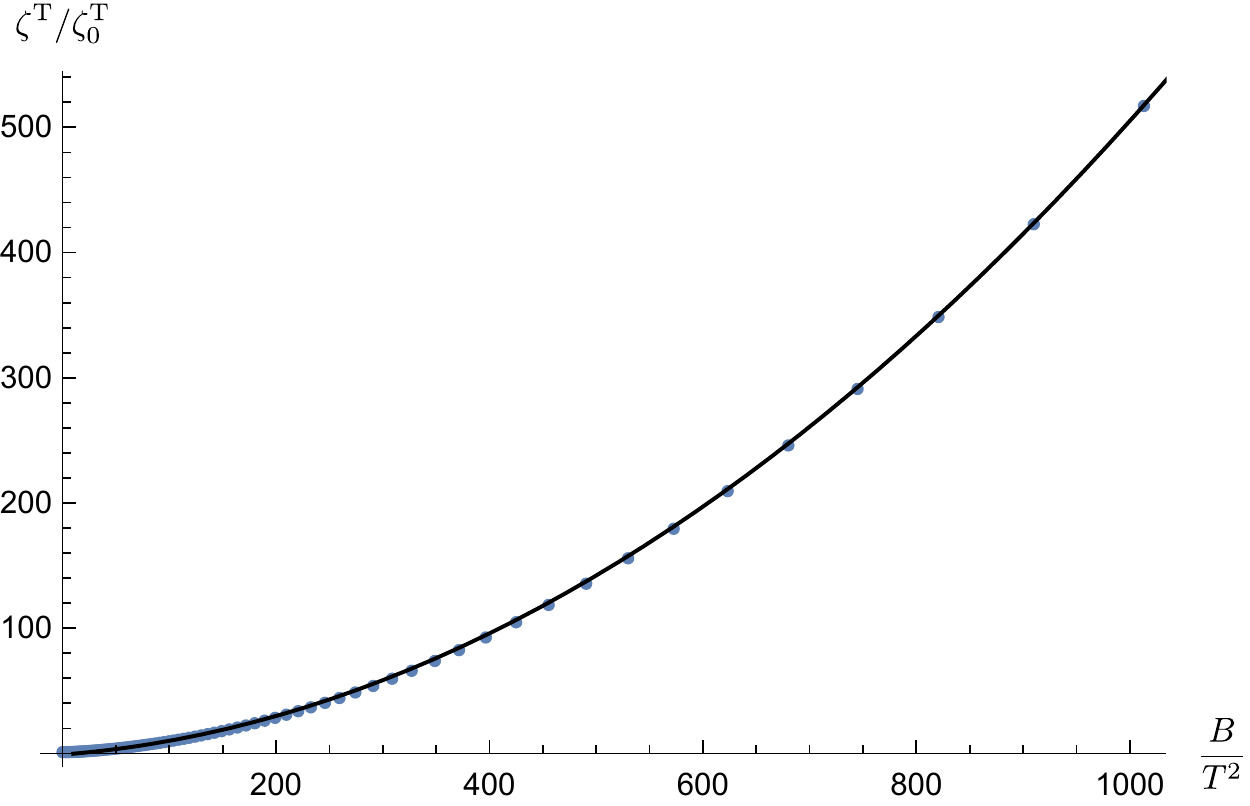}
	\includegraphics[width=0.48\textwidth]{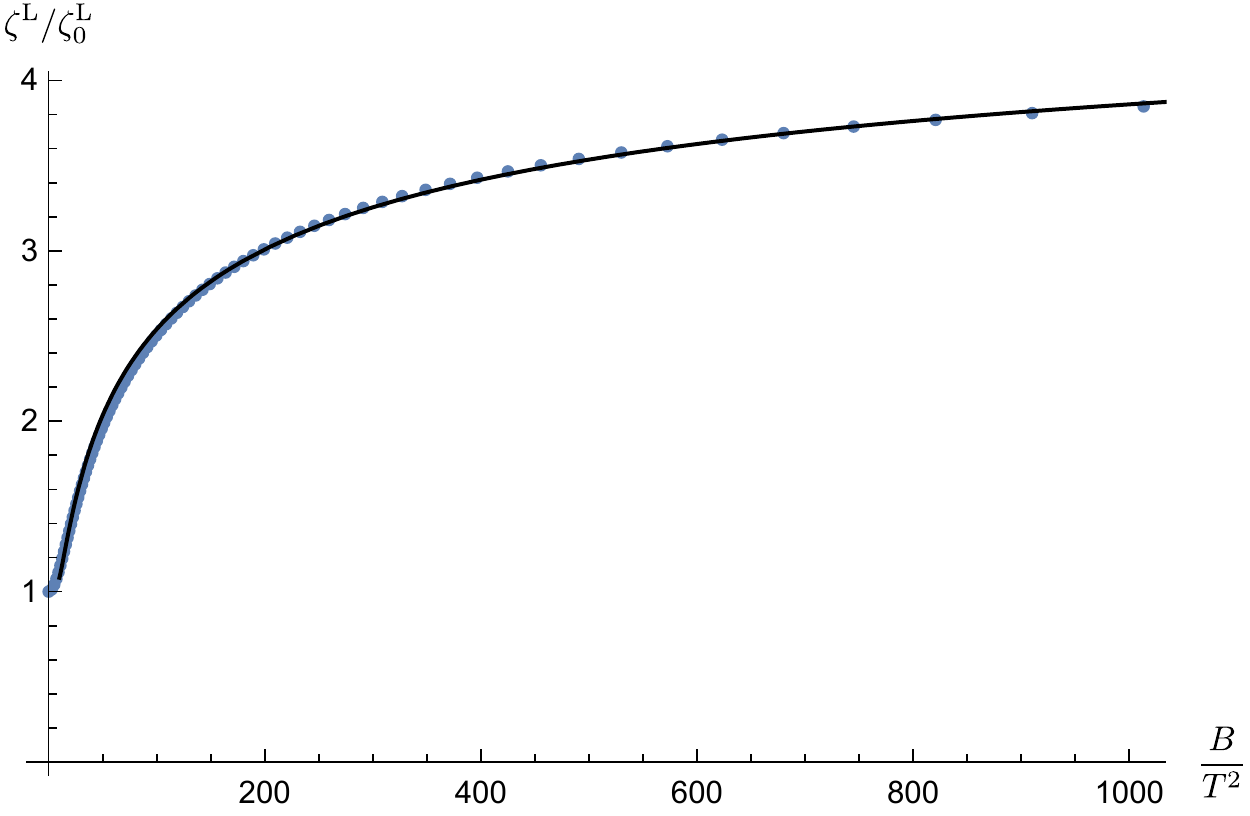}
	\includegraphics[width=0.48\textwidth]{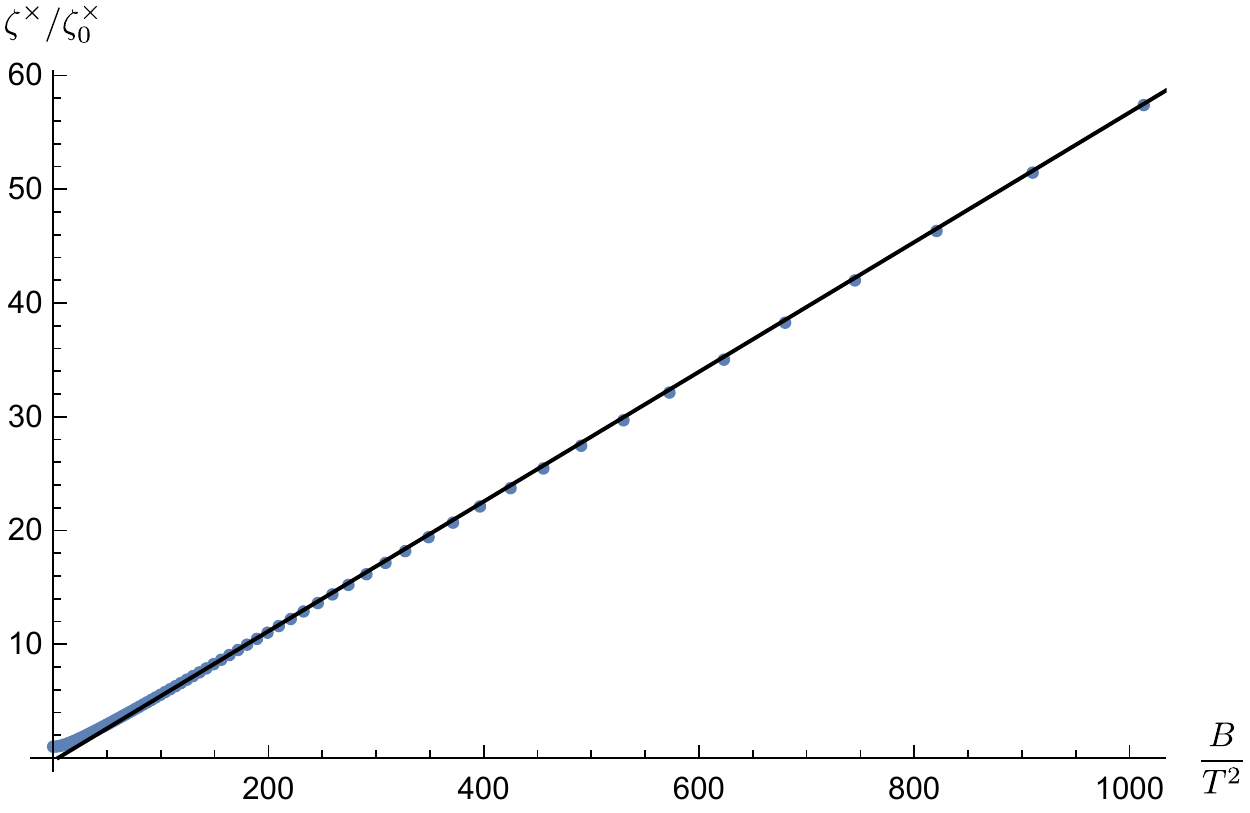}
	\caption{The numerical values (dots) of $\zeta^{\rm{T,L},\times}/\zeta^{\rm{T,L},\times}_{0}$ when $B/T^2$ is generic. The solid lines are fitting functions of \eqref{zeta_strongB}.} \label{figure_Grrrr}
\end{figure}

For generic value of magnetic field, we show numerical results for $\zeta^{\rm{T,L}}, \zeta^\times$ in Figure \ref{figure_Grrrr}. Obviously, all the coefficients grow as magnetic field is increased. In the strong magnetic field limit $B/T^2 \gg 1$, $\zeta^{\rm{T,L},\times}/ \zeta^{\rm{T,L},\times}_0$ are well fitted as
\begin{align}
&\frac{\zeta^{\rm T}}{\zeta^{\rm T}_0} \to -0.95 + 0.066 \frac{B}{T^2} + 0.00044 \left( \frac{B}{T^2} \right)^2, \qquad \frac{\zeta^{\rm L}}{\zeta^{\rm L}_0} \to 5.05 - 5.45 \frac{\log (B/T^2)}{\sqrt{B/T^2}}, \nonumber \\
&\frac{\zeta^\times}{\zeta^\times_0} \to -0.25 + 0.057 \frac{B}{T^2}, \label{zeta_strongB}
\end{align}
which hold for a reasonably wide range of $B/T^2$. The coefficient $\zeta^{\rm L}$ behaves similar as its quadratic counterpart $\eta_0^{\rm L}$, and shows a mild growth as $B/T^2$ is increased, and eventually saturates as $ \zeta^{\rm L}/\zeta^{\rm L}_0 \lesssim 5.05$. However, the coefficients $\zeta^{\rm T}, \zeta^\times$ increase more dramatically for strong magnetic field.

\section{Summary and Discussion} \label{summary}

From the perspective of action principle, we presented a comprehensive study on effective description of a Brownian particle moving in a magnetized plasma. First, within the framework of non-equilibrium EFT \cite{Crossley:2015evo,Glorioso:2017fpd,Glorioso:2018wxw} , we identify all the symmetries and construct effective action for Brownian particle, up to quartic order in particle's position. Then, we confirm the result through a model study based on holographic prescription for SK contour. Moreover, in the holographic model, we compute various coefficients in the effective action as functions of magnetic field and temperature, focusing on strong magnetic field limit.

Due to presence of non-Gaussian terms, it becomes inconvenient to cast the effective action into stochastic Langevin type equation \cite{Crossley:2015evo}. Nevertheless, we successfully convert the non-Gaussian effective action into a deterministic Fokker-Planck type equation, which corresponds to truncated Kramers-Moyal master equation at quartic order in derivatives. The Fokker-Planck type equation is more efficient for computing observables, such as moments of position/velocity of Brownian particle. It will be interesting to carry out a numerical study based on our non-Gaussian theory and clarify phenomenological consequences of non-Gaussian interactions \cite{Jana:2021niz}.

While dynamical KMS symmetry \eqref{dynamical_KMS_symmetry}-\eqref{dynamical_KMS_transform} is on the quantum level, the constant translational symmetry \eqref{constant_translation} renders the effective thoery to be entirely classical, in which quantum fluctuation is switched off. Relaxing the symmetry \eqref{constant_translation}, we have realized that, from non-equilibrium EFT perspective, the classical statistical limit \eqref{classical_statistical_limit} and quantum level \eqref{dynamical_KMS_transform} of dynamical KMS symmetry will give rise to different KMS relations among coefficients in the effective action. It will be interesting to investigate on this point via a direct holographic calculation, by considering an open string moving in a slowly-varying AdS black hole \cite{Abbasi:2012qz,Abbasi:2013mwa,Lekaveckas:2013lha,Reiten:2019fta} of fluid-gravity correspondence \cite{Bhattacharyya:2008jc}. Moreover, this new setup is supposed to yield more realistic \cite{Petrosyan:2021lqi} effective description for Brownian motion.

\appendix

\section{ Subtlety due to non-commutativity of $\epsilon \to 0$ versus $\omega \to 0$} \label{non-commutativity}

By adopting the method of \cite{Bu:2021jlp}, we now show that subtlety arising from non-commutativity \eqref{non-commute-issue} becomes accidentally irrelevant for the purpose of evaluating \eqref{I_integral1-4} up to $\mathcal{O}(\omega^1)$. With linearized string profile $X_{(1)}^{i,z}$ presented in \eqref{X_linearized_solution}, it is straightforward to show that contour integrals in \eqref{I_integral1-4} could be classified into three distinguished pieces
\begin{align}
\mathcal I_m = \mathcal I_{\rm ana} + \mathcal I_{\rm poles} + \mathcal I_{\rm non-ana}
\end{align}
Here, the first piece $\mathcal I_{\rm ana}$ vanishes since its integrand does not contain any singularity near the horizon. The second piece $\mathcal I_{\rm poles}$ could be simply computed by residue theorem as its integrand contains simple poles (no branch cuts) at the horizon. The last piece $\mathcal I_{\rm non-ana}$ involves logarithmic branch cut (maybe poles as well) at the horizon, which has a schematic form
\begin{equation}\label{}
\mathcal{I}_{\rm non-ana}=\int_{0_1}^{0_2} \frac{ du }{ u^2 }(u-1)^{\pm {\rm i} \hat \omega} \mathcal{H}(u,\hat \omega),
\end{equation}
where a potential factor $1/u^2$ is factorized, which would bring in UV divergence. Here, we use $\hat \omega$ to denote certain linear combination of $\omega_{1,2,3,4}$.  For generic value of $\hat \omega$, we do not have analytical expression for $\mathcal{H}(u,\hat \omega)$ for generic $\hat \omega$. However, we do know that $(u-1)^2\mathcal{H}(u,\hat \omega)$ is finite, non-singular and continuous inside the radial contour of Figure \ref{ucontour}. Thanks to the Weierstrass approximation theorem, it is legal to represent $(u-1)^2 \mathcal{H}$ by Taylor series $\sum_{l=-2}^{\infty}\mathcal{H}_l(\omega) (u-1)^{l+2}$ when $u$ is inside the radial contour. Thus,
\begin{equation}\label{WeiSeries}
\mathcal{I}_{\rm non-ana}=\sum_{l=-2}^{\infty} \mathcal{H}_l(\hat \omega) \mathfrak{I}_l, \qquad {\rm with} \quad  \mathfrak{I}_l \equiv \int_{0_1}^{0_2} \frac{ du }{ u^2 }(u-1)^{\pm {\rm i} \hat \omega} (u-1)^l.
\end{equation}
Therefore, the original task of computing \eqref{I_integral1-4} boils down to calculating simpler contour integrals $\mathfrak{I}_l$ of \eqref{WeiSeries}, which could be worked out analytically for generic value of $\hat \omega$. Afterwards, we extract low frequency limit of $\mathfrak{I}_l$ (see appendix B of \cite{Bu:2021jlp}):
\begin{align}
	& \mathfrak{I}_l = \mp \hat\omega  \frac{ _2F_1(2,n+1;n+2;1-\Lambda)}{(n+1) T} + \mathcal{O} (\hat\omega^2), \qquad \quad l \geq 1, \nonumber \\
	& \mathfrak{I}_l = \mp \frac{\hat\omega }{ \Lambda T}+\mathcal{O}(\hat\omega^2), \qquad \qquad \qquad \qquad \qquad \qquad ~~~~ l= 0, \nonumber \\
	&\mathfrak{I}_n = 2 {\rm i} \pi \mp \frac{ \hat\omega }{\Lambda T} + \mathcal{O} (\hat\omega^2), \qquad \qquad \qquad \qquad \qquad \quad \, l=-1, \nonumber \\
	&\mathfrak{I}_l = 4 {\rm i} \pi +\mathcal{O}(\hat\omega), \qquad \qquad \qquad \qquad \qquad \qquad \quad ~~~~ \,\, l=-2, \label{In_results}
\end{align}
where $\Lambda$ represents a UV cutoff near the AdS boundary $u=0$, and $_2F_1$ is a hypergeometric function. It is direct to check that the results \eqref{In_results} could be correctly recovered by first expanding the integrand of \eqref{WeiSeries} in small $\hat \omega$ and then computing the radial integral. However, this latter treatment cannot correctly cover higher order terms omitted in \eqref{In_results}, which correspond to higher derivative terms in the effective action. Therefore, in order to extracting order $\mathcal{O}(\omega^1)$ part of quartic effective action \eqref{I_quartic_holo}, it is valid to first expand the integrands (including the oscillating factor like $(u-1)^{\pm {\rm i} \hat \omega}$) in \eqref{I_integral1-4} in small $\hat \omega$, and then implement the radial integral.

\section{KMS relations when \eqref{constant_translation} is relaxed} \label{KMS_all}

In this appendix, we show that once the constant translational invariance \eqref{constant_translation} is relaxed, the classical statistical limit \eqref{classical_statistical_limit} and the high-temperature limit \eqref{hbar_beta_expansion} will give different KMS relations among coefficients in the effective action.

First, the quadratic Lagrangian \eqref{EFquadratic} receives corrections
\begin{align}
L_{\rm SK}^{(2),\, \rm new} = L_{\rm SK}^{(2)} +\theta_1 q^z_a q^z_r+\theta_2 q^i_a q^i_r+\theta_3B\epsilon_{ij}q^i_rq^j_a+\theta_4 B\epsilon_{ij} \dot q_a^i q_a^j
\end{align}
Imposing dynamical KMS symmetry \eqref{dynamical_KMS_symmetry} under classical statistical limit \eqref{classical_statistical_limit} and high-temperature limit \eqref{hbar_beta_expansion}, we find the same KMS relations
\begin{align}
	\eta_0^{\rm T} = \frac{1}{2} \beta \xi_0^{\rm T}, \qquad \eta^{\rm L}_0 = \frac{1}{2}\beta \xi_0^{\rm L}, \qquad \theta_3=0.
\end{align}

Next, we turn to corrections of the quartic Lagrangian \eqref{EFquartic}
\begin{align}
L_{\rm SK}^{(4), \, \rm new} = L_{\rm SK}^{(4)} + \delta L_{\rm SK}^{(4)}, 
\end{align}
where\footnote{For simplicity we have ignored terms containing an anti-symmetric tensor $\epsilon_{ij}$, which, under KMS transformation \eqref{dynamical_KMS_transform}, will not get interference with $\delta L_{\rm SK}^{(4)}$. Thus, inclusion of them will not modify the main conclusion.}, due to breaking of isotropy invariance, $\delta L_{\rm SK}^{(4)}$ looks lengthy,
\begin{align}
\delta L_{\rm SK}^{(4)} & = \lambda_{0,1} \dot q^i_a q^i_a (q^z_a)^2 +\kappa_{1,1} q^i_r q^i_a (q^j_a)^2+\kappa_{1,2} q^i_r q^i_a (q^z_a)^2+\kappa_{1,3} q^z_r q^z_a (q^i_a)^2+\kappa_{1,4} q^z_r (q^z_a)^3\nonumber\\
&+\lambda_{1,1}  q^i_r \dot q^i_a (q^j_a)^2 +\lambda_{1,2}  q^i_r \dot q^i_a (q^z_a)^2 +\lambda_{1,3}  q^z_r \dot q^z_a (q^i_a)^2  +\frac{\rm i}{2!} \left[\kappa_{2,1} (q^i_r)^2(q^j_a)^2+\kappa_{2,2} q^i_r q^i_a q^j_r q^j_a \right. \nonumber\\
& \left. +\kappa_{2,3} (q^i_r)^2 (q^z_a)^2+\kappa_{2,4} (q^z_r)^2 (q^i_a)^2+\kappa_{2,5} q^i_r q^i_a q^a_r q^z_a +\kappa_{2,6} (q^z_r)^2 (q^z_a)^2 \right] +\lambda_{2,1} \dot q^i_r q^i_r (q^j_a)^2 \nonumber\\
&  +\lambda_{2,2} \dot q^i_r q^i_a q^j_r q^j_a+\lambda_{2,3} \dot q^i_r q^i_r (q^z_a)^2+\lambda_{2,4} (q^i_a)^2 \dot q^z_r q^z_r +\lambda_{2,5} \dot q^i_r q^i_a q^z_r q^z_a+ \lambda_{2,6}  q^i_r \dot q^i_a q^z_r q^z_a \nonumber\\
& +\lambda_{2,7}  q^i_r q^i_a \dot q^z_r q^z_a+ \lambda_{2,8}  q^z_r \dot q^z_r (q^z_a)^2  +\kappa_{3,1} (q^i_r)^2 q^j_r q^j_a+\kappa_{3,2} q^i_a q^i_r (q^z_r)^2+\kappa_{3,3} q^z_r q^z_a (q^i_r)^2 \nonumber\\
&+\kappa_{3,4} q^z_r q^z_a (q^z_r)^2+\lambda_{3,1} (q^i_r)^2 \dot q^j_r q^j_a +\lambda_{3,2} \dot q^i_r q^i_r q^j_r q^j_a+\lambda_{3,3} q^i_a q^i_r (q^z_r)^2+\lambda_{3,4} q^i_a q^i_r \dot q^z_r q^z_r \nonumber\\
&+\lambda_{3,5} \dot q^z_r q^z_a (q^i_r)^2+\lambda_{3,6}q^z_r \dot q^z_a (q^i_r)^2+\lambda_{3,7} \dot q^z_r q^z_a (q^z_r)^2. \label{L_SK_4th_correction}
\end{align}
Imposing dynamical KMS symmetry \eqref{dynamical_KMS_symmetry} in the high-temperature limit \eqref{hbar_beta_expansion}, we find
\begin{align}
&\lambda_{0,1}= \frac{1}{8} i \left(\beta  \hbar^2 \kappa_{1,2}-\beta  \hbar^2 \kappa_{1,3}\right),\qquad \qquad \quad \, \kappa^{\rm T} = \frac{1}{48} \left(3 \beta \hbar^2 \kappa_{2,2}-4 \beta  \zeta^{T}\right), \nonumber \\
& \lambda_{1,1}= \frac{1}{16} \left(\beta \hbar^2 \kappa_{2,2}-2 \beta \hbar^2 \kappa_{2,1}\right), \qquad \qquad \,\,\, \kappa_2^{\times}= \frac{1}{96} \left(3 \beta \hbar^2 \kappa_{2,5}-4 \beta  \zeta^{\times}\right),\nonumber \\
&\lambda_{1,2}= \frac{1}{32} \left(\beta \hbar^2 \kappa_{2,5}-4 \beta \hbar^2 \kappa_{2,3}\right),\qquad \qquad \,\,\, \kappa_1^\times= \frac{1}{96} \left(3 \beta \hbar^2 \kappa_{2,5}-4 \beta  \zeta^{0}\right),\nonumber\\
&\lambda_{1,3}= \frac{1}{32} \left(\beta \hbar^2 \kappa_{25}-4 \beta \hbar^2 \kappa_{2,4}\right),\qquad \qquad \,\,\,\, \kappa^{\rm L} = \frac{1}{24} \left(\beta \hbar^2 \kappa_{2,6}-2 \beta  \zeta^{L}\right),\nonumber \\
&\lambda_{2,1}= \frac{\rm i}{8} \left(4 \beta  \kappa_{1,1}-\beta \hbar^2 \kappa_{3,1}\right), \qquad \qquad \qquad \lambda_{2,2}= \frac{\rm i}{4} \left(4 \beta  \kappa_{1,1}-\beta \hbar^2 \kappa_{3,1}\right),\nonumber \\
&\lambda_{2,3}= \frac{\rm i}{8} \left(4 \beta  \kappa_{1,2}-\beta \hbar^2 \kappa_{3,3}\right),\qquad \qquad \qquad \lambda_{2,4}= \frac{\rm i}{8} \left(4 \beta  \kappa_{1,3}-\beta \hbar^2 \kappa_{3,3}\right),\nonumber\\
&\lambda_{2,5}= \frac{\rm i}{4} \left(4 \beta  \kappa_{1,3}-\beta \hbar^2 \kappa_{3,3}\right),\qquad \qquad \qquad  \lambda_{2,7}= \frac{\rm i}{4} \left(4 \beta  \kappa_{1,2}-\beta \hbar^2 \kappa_{3,3}\right),\nonumber \\
&\lambda_{2,8}= \frac{3 \rm i}{8} \left(4 \beta  \kappa_{1,4}-\beta \hbar^2 \kappa_{3,4}\right),\qquad \qquad \qquad  \lambda_{2,6}= 0,\nonumber \\
&\lambda_{3,1}= -\frac{1}{2} \beta  \kappa_{2,1},\qquad \qquad \lambda_{3,2}= -\frac{1}{2} \beta  \kappa_{2,2}, \qquad \qquad \lambda_{3,3}= -\frac{1}{2} \beta  \kappa_{2,4},\nonumber \\
&\lambda_{3,4}= -\frac{1}{4} \beta  \kappa_{2,5}, \qquad \qquad \lambda_{3,5}= \frac{1}{8} (\beta  \kappa_{2,5}-4 \beta  \kappa_{2,3}), \qquad \lambda_{3,6}= \frac{1}{8}\beta  \kappa_{2,5},\nonumber \\
&\lambda_{3,7}= -\frac{1}{2} \beta  \kappa_{2,6}, \qquad \qquad \kappa_{3,2}= \kappa_{3,3}.  \label{KMS_high_temperature}
\end{align}
On the other hand, if we impose dynamical KMS symmetry \eqref{dynamical_KMS_symmetry} in the classical statistical limit \eqref{classical_statistical_limit}, we would get
\begin{align}
&\lambda_{0,1}=\lambda_{1,1}=\lambda_{1,2}=\lambda_{1,3}= \lambda_{2,6}=0,\nonumber\\
&\kappa^{\rm T} = -\frac{ 1 }{ 12 }\beta \zeta^{\rm T},\qquad \kappa_1^\times = \kappa_2^\times = -\frac{ 1 }{ 24 }\beta \zeta^{\rm \times}, \qquad \kappa^{\rm L} = -\frac{ 1 }{ 12 }\beta \zeta^{\rm L} \nonumber\\
&\lambda_{2,1}= \frac{\rm i }{ 2}\beta  \kappa_{1,1},\qquad \quad \lambda_{2,2}= {\rm i} \beta  \kappa_{1,1},\qquad \lambda_{2,3}= \frac{\rm i }{2 }\beta  \kappa_{1,2},\qquad \lambda_{2,4}=\frac{\rm i }{2 } \beta  \kappa_{1,3},\nonumber\\
&\lambda_{2,5}= {\rm i} \beta  \kappa_{1,3},\qquad \quad \lambda_{2,7}= {\rm i} \beta  \kappa_{1,2}, \qquad\lambda_{2,8}= \frac{3 \rm i }{ 2} \beta  \kappa_{1,4}, \qquad\lambda_{3,1}= -\frac{1}{2} \beta  \kappa_{2,1},\nonumber\\
&\lambda_{3,2}= -\frac{1}{2} \beta  \kappa_{2,2},\qquad \lambda_{3,3}= -\frac{1}{2} \beta  \kappa_{2,4},\qquad \lambda_{3,4}= -\frac{1}{4} \beta  \kappa_{2,5},\qquad \kappa_{3,2}=\kappa_{3,3},\nonumber\\
&\lambda_{3,5}= \frac{1}{8} (\beta  \kappa_{2,5}-4 \beta  \kappa_{2,3}), \qquad\lambda_{3,6}= \frac{1}{8}\beta  \kappa_{2,5},\qquad \lambda_{3,7}= -\frac{1}{2} \beta  \kappa_{2,6}, \label{KMS_classical}
\end{align}
which is actually $\hbar \to 0$ limit of \eqref{KMS_high_temperature}.

Obviously, if we require constant translational invariance \eqref{constant_translation}, say setting all the coefficients in \eqref{L_SK_4th_correction} to be zero, we will immediately see that the KMS relations \eqref{KMS_high_temperature} and \eqref{KMS_classical} will collapse to \eqref{quarticKMS}.

\section*{Acknowledgements}

We would like to thank Gao-Liang Zhou for helpful discussions.

\bibliographystyle{utphys}

\bibliography{reference}

\end{document}